\documentclass[journal=ancac3,manuscript=article,layout=twocolumn]{achemso}
\usepackage[version=3]{mhchem} % Formula subscripts using \ce{}
\usepackage{braket}
\usepackage{siunitx} 
\usepackage{color}
\usepackage{bm}
\author{Ryuta Tsukahara}
\affiliation[KGU]{School of Science and Technology, Kwansei Gakuin University, Sanda-shi, Hyogo, 669-1337, Japan}
\altaffiliation{Contributed equally to this work}
\author{Masazumi Fujiwara}
\affiliation[OCU]{Department of Chemistry, Osaka City University, Sumiyoshi-ku, Osaka, 558-8585, Japan}
\alsoaffiliation[KGU]{School of Science and Technology, Kwansei Gakuin University, Sanda-shi, Hyogo, 669-1337, Japan}
\altaffiliation{Contributed equally to this work}
\email{masazumi@osaka-cu.ac.jp}
\author{Yoshihiko Sera}
\affiliation[KGU]{School of Science and Technology, Kwansei Gakuin University, Sanda-shi, Hyogo, 669-1337, Japan}
\author{Yushi Nishimura}
\affiliation[OCU]{Department of Chemistry, Osaka City University, Sumiyoshi-ku, Osaka, 558-8585, Japan}
\author{Yuko Sugai}
\affiliation[KGU]{School of Science and Technology, Kwansei Gakuin University, Sanda-shi, Hyogo, 669-1337, Japan}
\author{Christian Jentgens}
\affiliation[Mic]{Microdiamant AG, Kreuzlingerstrasse 1, CH-8574 Lengwil, Switzerland}
\author{Yoshio Teki}
\affiliation[OCU]{Department of Chemistry, Osaka City University, Sumiyoshi-ku, Osaka, 558-8585, Japan}
\author{Hideki Hashimoto}
\affiliation[KGU]{School of Science and Technology, Kwansei Gakuin University, Sanda-shi, Hyogo, 669-1337, Japan}
\author{Shinichi Shikata}
\affiliation[KGU]{School of Science and Technology, Kwansei Gakuin University, Sanda-shi, Hyogo, 669-1337, Japan}
\email{SShikata@kwansei.ac.jp}
\title[An \textsf{achemso} demo]
  {Effect of surface oxidation on electron spin coherence of single nitrogen-vacancy centers in nanodiamonds }

\begin{document}


%%%%%%%%%%%%%%%%%%%%%%%%%%%%%%%%%%%%%%%%%%%%%%%%%%%%%%%%%%%%%%%%%%%%%%%
%\begin{tocentry}

%Some journals require a graphical entry for the Table of Contents.
%This should be laid out ``print ready'' so that the sizing of the

%Inside the \texttt{tocentry} environment, the font used is Helvetica
%8\,pt, as required by \emph{Journal of the American Chemical

%The surrounding frame is 9\,cm by 3.5\,cm, which is the maximum
%permitted for  \emph{Journal of the American Chemical Society}
%graphical table of content entries. The box will not resize if the
%content is too big: instead it will overflow the edge of the box.

%This box and the associated title will always be printed on a
%separate page at the end of the document.

%\end{tocentry}
%%%%%%%%%%%%%%%%%%%%%%%%%%%%%%%%%%%%%%%%%%%%%%%%%%%%%%%%%%%%%%%%%%%%%%

\begin{abstract}
Fluorescent nanodiamonds (NDs) are crystal-defect-based light-emitting nanoparticles that can be applied to quantum information science and quantum sensing. 
Of particular interest are nitrogen vacancy (NV) centers that allow optical access to their coherently controllable electron spin systems, leading to spin-photon quantum devices and  nanoscale sensing of various physical parameters.
However, the NV spin coherence time (${\rm T_2}$) in NDs has been limited to one or two orders of magnitude shorter than those in bulk diamonds owing to the complicated surface effect that decoheres the NV spin systems. 
Here, we study the relation between the surface properties and ${\rm T_2}$ of single NV centers in NDs by systematically analyzing the effect of surface oxidation. 
We apply aerobic and acidic oxidation methods with various heating temperatures and processing times and find that aerobic oxidation most effectively oxidizes the surface and extends ${\rm T_2}$ by a factor of 1.6 to the original NDs.
The ND-size dependence of ${\rm T_2}$ clearly shows that the surface oxidation removes a constant decoherence contribution irrespective of the ND size and that a new surface-derived decoherence source emerges when the ND size reduces below 50 nm. 
The present results provide quantitative information on the decoherence sources of NV spin systems of NDs and will enable a strategic surface modification for better spin manipulations of NV centers in the context of quantum information science and nanoscale quantum sensing.
\end{abstract}

\section{Introduction}
Fluorescent nanodiamonds (NDs) are crystal-defect-based light-emitting nanoparticles that can be used for applications in quantum information science and quantum nanoscale sensing. 
Of particular interest are nitrogen vacancy (NV) centers, which allow optical access to coherently controllable electron spin systems, leading to spin-photon quantum devices~\cite{togan2010quantum,kosaka2015entangled,almokhtar2014numerical,fujiwara2015ultrathin,schroder2012nanodiamond,fujiwara2016manipulation} and nanoscale sensing of various physical parameters.~\cite{doherty2013nitrogen,schirhagl2014nitrogen,iwasaki2017direct,neumann2013high,andrich2017long,PhysRevLett.119.240401}
The electron spin properties of NV centers are the key to a better performance of these applications. 
In particular, the electron spin coherence time (${\rm T_2}$) limits the spin memory time in quantum devices and the sensitivity of quantum sensing.
There has been a significant effort to extend ${\rm T_2}$ in both bulk diamonds and NDs.~\cite{rabeau2007single,wang2016coherence,knowles2014observing,tisler2009fluorescence,ohashi2013negatively,trusheim2013scalable,song2014statistical,laraoui2012nitrogen,boudou2013fluorescent} 
However, ${\rm T_2}$ in NDs is one or two orders of magnitude shorter than those in bulk diamonds because of the complicated surface effects, whose exact decoherence mechanism has not been clarified.~\cite{tisler2009fluorescence,song2014statistical,mcguinness2013ambient,chou2017nitrogen}
Thus, it is important to understand the surface effects in more detail to achieve an extended ${\rm T_2}$ in NDs.

Surface effects are a composite of various decoherence sources and have not been fully understood, particularly in NDs, where surface inhomogeneity exists.
In case of bulk diamonds, surface oxidation effectively extends ${\rm T_2}$ of NV centers.; 
Ohashi et al. reported that oxygen termination by using a strong acid mixture significantly extends ${\rm T_2}$ of NV centers located near the surface of high-purity bulk diamonds compared to hydrogen or fluorine termination.~\cite{ohashi2013negatively} 
Surface oxidation has been historically explored for NDs to clean the surface by removing sp$^2$-like carbon soot generated during the milling process.
Strong acid mixtures were used because the whole process was completed in the wet process, which is important for preserving a uniform dispersion of NDs.~\cite{tisler2009fluorescence,wolcott2014surface,trusheim2013scalable,lai2011modeling,nguyen2007adsorption}
Later, aerobic oxidation was applied as a more efficient oxidizing technique for NDs.~\cite{osswald2006control}
While acidic oxidation is enough to oxidize the surface of bulk diamonds in terms of ${\rm T_2}$ extension, aerobic oxidation seems necessary for NDs to completely oxidize the surface and extend ${\rm T_2}$.\cite{bradac2018effect,wolcott2014surface,gaebel2012size,stehlik2015size}

The importance of aerobic oxidation is related to the surface inhomogeneity that NDs comprehend. For example, during the milling process of original bulk diamonds, the ND surface gets covered with a significant amount of carbon soot; such sp$^2$-like materials turn the ND powder color from yellow to black.~\cite{wolcott2014surface} 
In addition, the surface termination of NDs cannot be defined simply as it can be for bulk diamonds, where X-ray photo-emission spectroscopy (XPS) determines even the surface coverage ratio of various oxygen terminals such as ester, carbonyl, and carboxyl.\cite{wang2011higher,fang2017synthesis} 
This surface inhomogeneity must be quantitatively understood to extend ${\rm T_2}$ in NDs. 
Statistical approaches examining the number of ND particles are required for obtaining the clear surface-${\rm T_2}$ relation undermined by surface inhomogeneity.
%~\cite{knowles2014observing,ermakova2013detection,trusheim2013scalable,song2014statistical,cao2017protecting,laraoui2012nitrogen,rios2010quantum,maze2008nanoscale,boudou2013fluorescent,tisler2009fluorescence,arnault2017nanodiamonds} 
It is important to quantitatively determine the surface-${\rm T_2}$ relation of NDs, particularly in relation to surface oxidation.

Here, we quantitatively study the effect of oxidation on ${\rm T_2}$ of single NV centers in NDs. 
Aerobic and acidic oxidations with various processing parameters are applied to NDs and their surfaces are characterized by surface-sensitive spectroscopies. 
By measuring the statistical data of ${\rm T_2}$ for the respective ND samples, 
we find that aerobic oxidation most effectively oxidizes the surface and extends ${\rm T_2}$ by a factor of 1.6 to the original NDs.
Moreover, the ND-size dependence of ${\rm T_2}$ for the aerobically oxidized NDs clearly shows that 
the surface oxidation removes a constant decoherence contribution irrespective of the ND size and that a new surface-derived decoherence source emerges when the ND size reduces below 50 nm. 
The present results provide quantitative information on the decoherence sources of the NV spin systems of NDs and will enable a strategic surface modification of NDs toward better spin manipulations of NV centers in the context of quantum information science and nanoscale quantum sensing.

\section{Results and Discussion}
\subsection{Surface oxidation and etching of NDs}

\begin{figure*}[t!]
\centering
  \includegraphics[scale=1.1]{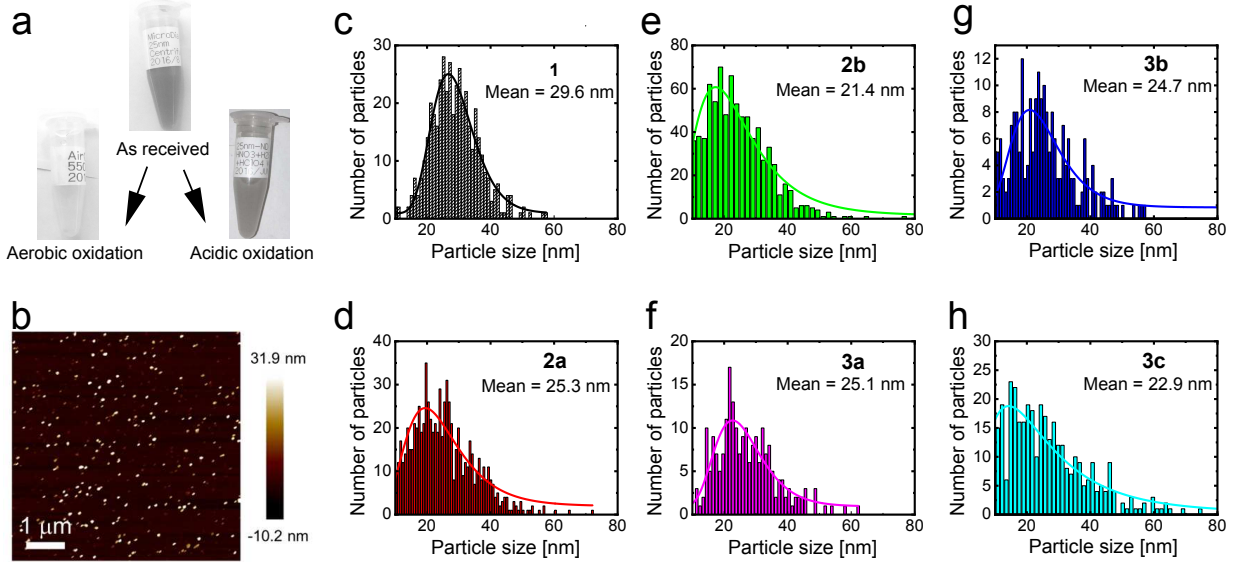}
  \caption{Surface oxidation and etching of NDs. (a) Pictures of ND suspensions after aerobic or acidic oxidation. (b) A topographic image of the as-received NDs on a coverslip. Particle-size-distribution histograms of (c) as-received ({\bf 1}), (d) acidic oxidation for 6 h ({\bf 2a}), (e) acidic oxidation for 24 h ({\bf 2b}), (f) aerobic oxidation at 450 $\si{\degreeCelsius}$ for 1 h ({\bf 3a}), (g) aerobic oxidation at 550 $\si{\degreeCelsius}$ for 1 h ({\bf 3b}), and (h) aerobic oxidation at 550 $^\circ$C for 2 h ({\bf 3c}).
  The solid lines indicate the log-normal fitting to the data. The mean particle sizes based on the fitting are shown in each figure.}
  \label{fgr:figure1}
\end{figure*}

We prepare multiple sets of NDs oxidized by aerobic or acidic oxidation with different temperatures and processing times (Fig.~\ref{fgr:figure1}a and see Methods). 
These NDs are spin-coated on cleaned coverslips for the size characterization using an atomic force microscope (AFM) (Fig.~\ref{fgr:figure1}b and Fig.~S1). 
Figures~\ref{fgr:figure1}c-h show their particle size distributions; as-received (hereafter designated as {\bf 1}), acidic oxidation for 6 and 24 h ({\bf 2a}, {\bf 2b}), and aerobic oxidation at 450$\si{\degreeCelsius}$ for 1 h, aerobic oxidation at 550 $\si{\degreeCelsius}$ for 1 h and 2 h ({\bf 3a}, {\bf 3b}, {\bf 3c}), respectively. 
The log-normal fitting gives the mean particle sizes of 29.6, 25.3, 21.4, 25.1, 24.7, and 22.9 nm, respectively, for Figs.~\ref{fgr:figure1}c--h (see Table S1 for the detail). 
The ND size decreases as the processing time increases both in the aerobic and acidic oxidations. 
For aerobic oxidation, 550 $\si{\degreeCelsius}$ more rapidly etches the NDs than 450 $\si{\degreeCelsius}$, which is in accordance with previous reports ({\bf 3a}, {\bf 3c})~\cite{osswald2006control,wolcott2014surface,wang2016coherence}. 

The etching rates obtained by aerobic oxidation do not show a significant differences among the processing temperature, 4.5, 4.9, and 3.4 nm/h for 
{\bf 3a}, {\bf 3b}, and {\bf 3c}, respectively, which are consistent with previous reports on ND oxidation (4 $\pm$ 1 nm/h at 550 $\si{\degreeCelsius}$)~\cite{gaebel2012size}. 
Acidic oxidation also etches the NDs with an etching rate of 0.7 nm/h and 0.3 nm/h for 6 and 24 h ({\bf 2a}, {\bf 2b}), respectively.
We confirmed that too intense oxidation, such as aerobic oxidation at 550 $\si{\degreeCelsius}$ for 6 h or acidic oxidation for 72 h, significantly reduced the number of NV centers in the following optically detected magnetic resonance (ODMR) experiments owing to hard etching~\cite{gaebel2012size,wang2016coherence}. 
Therefore, we focus on the oxidation range in which we can find enough number of NV centers for obtaining the statistical values of ${\rm T_2}$ in this study.  

%酸処理のエッチングレート修正・追加しました。

%The etching rate is dependent on the particle size; the smaller the nanodiamonds are, the etching rate is faster,~\cite{bradac2018effect} while we have not observed the significant difference of the etching rates 

\subsection{Surface characterization}
 \begin{figure*}[t!]
 \centering
 \includegraphics[scale=1.2]{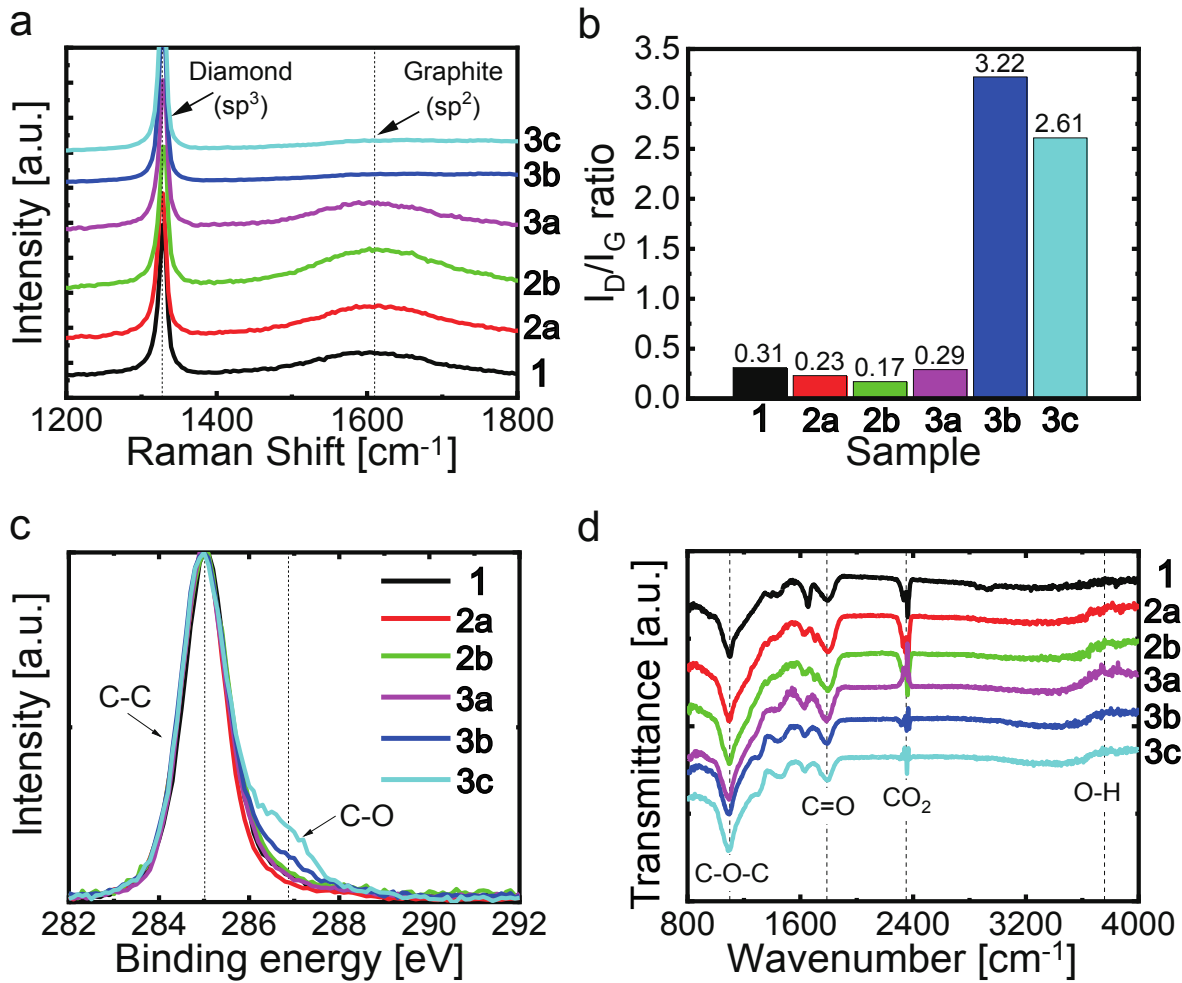}
 \caption{Surface characterization of ND samples using various spectroscopies. (a) Raman spectra (325 nm excitation) and (b) the peak ratio of the diamond line and the surface graphitic band ($I_D/I_G$), where the peak area is regarded as their intensities. (c) XPS spectra of the NDs near C1s transitions. (d) FTIR spectra of the NDs.}
 \label{fgr:figure2}
\end{figure*}

We next characterize the surface properties of these ND samples using Raman spectroscopy, XPS, and Fourier-transform infrared spectroscopy (FTIR), which can provide detailed information of the surface.
Figure~\ref{fgr:figure2}a shows Raman spectra of the ND samples excited at 325 nm, which is resonant to band-to-band transition.
A sharp peak observed at 1333 $\si{cm}^{-1}$ is the symmetric stretching mode of the C-C bond of the diamond lattice structure (so-called diamond line) and a broad peak observed at around 1600 $\si{cm}^{-1}$ is the G band of the surface graphitic layers (sp$^2$-like) of ND particles ~\cite{wolcott2014surface,popov2017raman,bradac2018effect,mochalin2008contribution,rondin2010surface,laube2018photo}. 
The intensity ratio of the diamond line and G band ($I_D/I_G$) is shown in Fig.~\ref{fgr:figure2}b. 
The aerobic oxidation at 550 $\si{\degreeCelsius}$ decreases the G band and intensifies the diamond line ({\bf 3b}, {\bf 3c}), resulting in an increase in $I_D/I_G$. 
At 550 $\si{\degreeCelsius}$, the G band is completely removed already for the duration of 1 h ({\bf 3b}) and no significant change is observed for the further oxidation up to 2 h ({\bf 3c}) ($I_D/I_G$ ratio seems to decrease within an error). 
The heating temperature of 450 $\si{\degreeCelsius}$ has a weak effect and the G band remains ({\bf 3a}). 

In contrast, the acidic oxidation does not seem to remove the graphitic layers with the present temperature of 80 $\si{\degreeCelsius}$ ({\bf 2a}, {\bf 2b}), while the ND size is decreased, as evidenced in the ND size measurements (see Figs.~\ref{fgr:figure1}g, h). 
The processing times of 6 and 24 h for acidic oxidation do not show significant differences in the Raman spectra, while the ND size is substantially decreased for this duration. %color turns slightly pale gray from black. 
Further extension of the processing time etches the surface more, which however results in further size reduction of NDs eliminating the presence of NV centers.
In fact, we cannot find NV centers for the duration of 72 h in acidic oxidation.
These results clearly show the effectiveness of aerobic oxidation in removing the surface graphitic layers of NDs.

Figure~\ref{fgr:figure2}c shows XPS spectra of the ND samples.　
The prominent peak observed at 284.8 eV is ascribed to 1s of sp$^3$ C-C bond and observed in all NDs~\cite{laube2017defined}. 
The NDs with aerobic oxidation at 550 $\si{\degreeCelsius}$ ({\bf 3b}, {\bf 3c}) have peak tails in the higher energy side in the region of 286--288 eV. 
These tails are convolutions of the oxygen-related bonding peaks such as ester (C-O), carbonyl (C=O), and carboxyl (COOH);~\cite{wang2011higher,inel2016solvent} however, the energy difference of these functional groups cannot be detected owing to surface inhomogeneity and we cannot obtain detailed information of surface termination from the XPS data only, which is a striking contrast to bulk diamonds, where even the surface coverage ratio of the functional groups is determined through XPS.

Figure~\ref{fgr:figure2}d shows the FTIR spectra (see Supporting Information for the detailed assignment). 
A strong peak is observed in the range of 1000--1300 cm$^{-1}$, which is ascribed to -C-O-C- stretching vibrations of cyclic ethers~\cite{jiang1996ftir}. 
The peaks at 1307 and 1451 cm$^{-1}$ are attributed to CO bending vibration and asymmetric -CH bending vibration, respectively~\cite{shenderova2011hydroxylated} ({\bf 2a}, {\bf 2b}).
A weak peak appearing at 1627 cm$^{-1}$ in the NDs with acidic oxidation is owing to the -OH bending vibration. 
This -OH bending either comes from the carboxyl group (-COOH) of the ND surface or water molecules adsorbed on the sample surface.~\cite{mochalin2008contribution} 
A peak at 1782 cm$^{-1}$ results from the C=O stretching mode and is attributed to the presence of carboxyl.~\cite{gines2017positive} 

These results of the surface-sensitive spectroscopies show that aerobatic oxidation effectively removes the graphitic layers on the ND surface and performs surface termination with oxygen, while the surface inhomogeneity of NDs prevents the formation of uniform and well-defined surfaces that one can obtain in bulk diamonds.~\cite{wang2011higher} 
%Note that 550 $\si{\degreeCelsius}$ is relatively higher than the temperature generally used for surface oxidation.~\cite{osswald2006control}
%However, we use this temperature to apply hard-etching to confirm the complete removal of the graphitic layers by referring to a previous report.~\cite{wolcott2014surface}
%The heating conditions are also strongly dependent on the heating procedures, such as the pre-heating time and the real temperature of the samples, and should have uncertainty between the reports. 
%We therefore think it is not straightforward to compare the detailed heating parameters with the previous reports

While the aerobic oxidation already exhibited the effectiveness to NDs, the effect of acidic oxidation is not clear at present. 
Acidic oxidation significantly etches the ND particles (reduces the ND size), as seen in Fig.~\ref{fgr:figure1}, however,
the surface graphitic layers have not been removed well.
The spin coherence of NV centers is also not prolonged by acidic oxidation, as we will see in the following section.
Because the etching rate of acidic oxidation is significantly smaller than that of aerobic oxidation and it seems to slow down (0.7 to 0.3 nm/h for 6 to 24 h), the acidic oxidation in the present experiment does not seem to be simply etching the surface and graphitic layers. 
The detailed chemical-reaction process of ND surface oxidation has been only recently investigated~\cite{bradac2018effect} and the processing parameters of the present acidic oxidation needs to be studied in more detail. 
%Nevertheless our present study can still obtain a quantitative information on the relation between the NV spin properties and the surface properties, which we show in the following sections.
%Note that we have not been able to optimize the acidic oxidation parameters.
For example, a much higher temperature, such as 200 $\si{\degreeCelsius}$, is generally used for bulk diamonds and can be effective to the surface oxidation of NDs.~\cite{hauf2011chemical,fu2010conversion,ohno2012engineering} 
However, it is impossible to apply such high temperatures to the present small quantity of ND samples because most of the NDs (and hence NVs) become extinct at this temperature. 
One may have to use a large portion of ND suspension for such high-temperature acidic oxidation to obtain a substantial amount of NDs for the surface characterization or ODMR measurements.

\begin{figure*}[t!]
 \centering
 \includegraphics[scale=1.0]{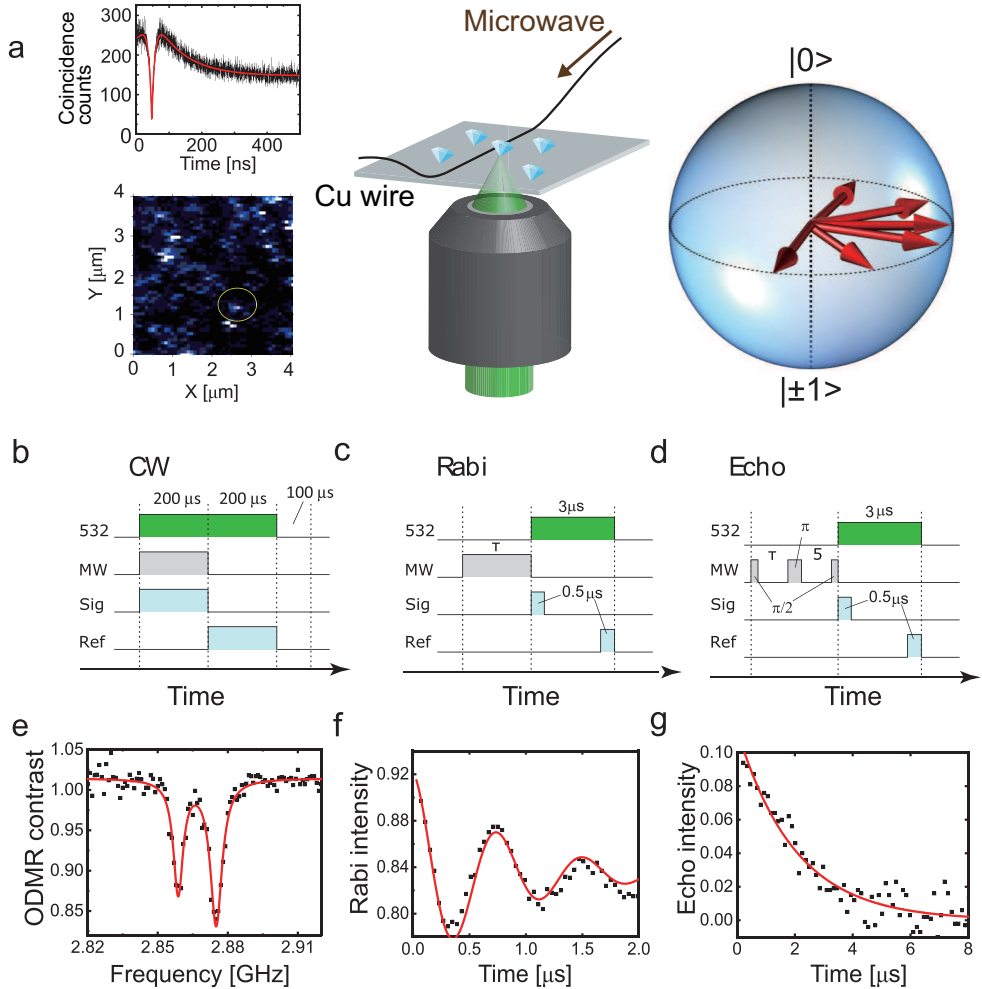}
 \caption{(a) Close-up of the central part of the experimental setup and schematic of the NV-spin precession during spin-echo sequence. The NDs are placed on a coverslip. A thin copper wire is used as a linear microwave antenna for spin excitation. Single NV centers hosted in NDs are excited by 532-nm laser light and observed with red-shifted fluorescence collected through the same microscope objective. A representative confocal fluorescence scanning image and second-order photon correlation histogram of single NV centers are shown as insets. Schematic illustrations of the gated photon counting for (b) CW, (c) Rabi, and (d) spin-echo measurements.
 532: 532-nm green laser pulse. MW: microwave pulse. Sig: photon counting while the microwave is ON. Ref: photon counting while the microwave is OFF. (e) Typical CW-ODMR profile of single NV centers in NDs and its (f) Rabi and (g) spin-echo profiles. 
}
 \label{fgr:figureExp}
\end{figure*}

\begin{figure}[b!]
 \centering
 \includegraphics[scale=1.6]{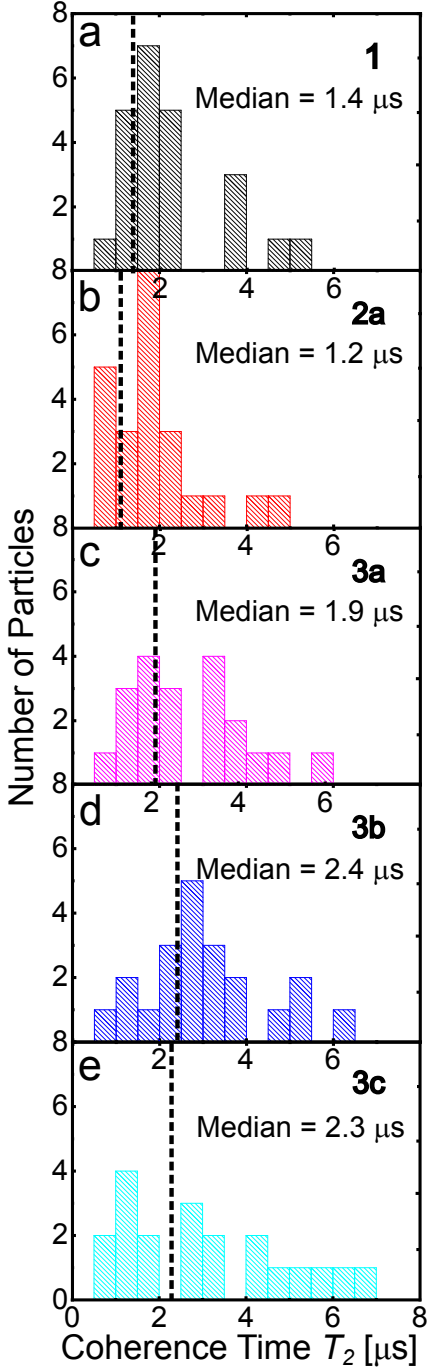}
 \caption{Statistical data of the spin coherence time (${\rm T_2}$) for different NDs: (a) as-received ({\bf 1}), (b) acidic oxidation for 6 h ({\bf 2a}), (c) aerobic oxidation at 550 $\si{\degreeCelsius}$ for 1 h ({\bf 3a}), (d) aerobic oxidation at 550 $\si{\degreeCelsius}$ for 2 h ({\bf 3b}), and (e) aerobic oxidation at 450 $\si{\degreeCelsius}$ for 1 h ({\bf 3c}). The dashed lines indicate the median values. The number of particles investigated is 22, 23, 20, 21, and 20 for {\bf 1}--{\bf 3c}, respectively. }
 \label{fgr:figure3}
\end{figure}

\begin{table*}[th!]
\small
  \caption{\ ${\rm T_2}$ of each ND sample with the maximum and median.}
  \label{tbl:table1}
 %\scalebox{0.68}[1]{
 \begin{tabular}{cccccc}
   \hline
    ND sample & {\bf 1} & {\bf 2a} & {\bf 3a} & {\bf 3b} & {\bf 3c} \\
    \hline
    Maximum [$\si{\us}$] & 4.7 & 4.2 & 5.4 & 5.5 & 6.1 \\
    Median [$\si{\us}$] & 1.4 & 1.2 & 1.9 & 2.4 & 2.3 \\
    \hline
 \end{tabular}
\end{table*}

\subsection{Effect of surface oxidation on ${\rm T_2}$}

The electron spin properties of these ND samples are measured with a home-built confocal fluorescence scanning microscope equipped with a microwave excitation system for spin characterization (see Methods and Supporting Information).~\cite{fujiwara2018tracking,fujiwara2018observation}
The NV centers we find in the present NDs are mostly single owing to the low NV density. 
By choosing only single NV centers out for the ODMR experiments on the basis of antibunching detection in the second-order photon correlation histogram (Fig.~\ref{fgr:figureExp}a), 
we determine ${\rm T_2}$ of multiple ND particles (N $\geq$ 20) by sequentially applying CW, Rabi, and spin-echo measurements to either of the spin transitions of $\ket{0} \rightarrow \ket{\pm 1}$, as explained in Figs.~\ref{fgr:figureExp}b-e. 
A weak external magnetic field of $\sim$ 30 Gauss is applied to separate the two ODMR peaks sufficiently to prevent the mutual interference of the split peaks in the pulsed ODMR measurements. 
The measured spin-echo profile is fitted with a single exponential decay to determine ${\rm T_2}$ (Fig.~\ref{fgr:figureExp}g). 
The approximation of the decay profile as a single exponential is validated for short coherence time, as described  elsewhere.~\cite{tisler2009fluorescence,knowles2014observing,mcguinness2013ambient}
The obtained data sets for the number of NDs enable us to quantitatively study the relation between  ${\rm T_2}$ and the surface oxidation covered under the ND's intrinsic surface inhomogeneity. 

Figures~\ref{fgr:figure3}a-e show statistical histograms of ${\rm T_2}$ of the ND samples, {\bf 1}, {\bf 2a}, {\bf 3a}, {\bf 3b}, and {\bf 3c}, respectively.
Table~\ref{tbl:table1} summarizes the statistical figures (median and maximum). 
The as-received NDs ({\bf 1}) show the median ${\rm T_2}$ of 1.4 $\si{\us}$ with a longest ${\rm T_2}$ of 4.7 $\si{\us}$. 
The aerobically oxidized NDs show the medians of 1.9 $\si{\us}$, 2.4 $\si{\us}$, and 2.3 $\si{\us}$ for the ND samples, {\bf 3a}, {\bf 3b}, and {\bf 3c}, respectively.
The longest ${\rm T_2}$ for aerobic oxidation is found in {\bf 3c}. 
The median ${\rm T_2}$ is increased by a factor of 1.71 for the 1-h aerobic oxidation ({\bf 3b}) and a factor of 1.64 for the 2-h aerobic oxidation ({\bf 3c}), compared to the as-received NDs ({\bf 1}). 
In contrast, the {\bf 2a} NDs do not show any changes in ${\rm T_2}$ compared to the {\bf 1} NDs, indicating the ineffectiveness of acidic oxidation with the present processing parameters.
These results show that surface oxidation is almost completed for an hour of the aerobic oxidation at 550$\si{\degreeCelsius}$ and the 6-h acidic oxidation does not increase ${\rm T_2}$ significantly, which is consistent with the above surface characterization data.

Note that the measured ${\rm T_2}$ depends on the angle between the NV axis and the microwave magnetic field polarization, whereas the present experimental setup does not compensate this angle dependence because of the random NV orientations of the NDs on the coverslip.~\cite{stanwix2010coherence,maze2008electron} 
Therefore, the present results are the net distribution of ${\rm T_2}$ including the uncertainty of the orientation differences. 
However, the angle dependence equally influences all samples and does not affect the following discussion.

\subsection{Detailed analysis of decoherence sources of NDs}
\begin{figure}[b!]
 \centering
 \includegraphics{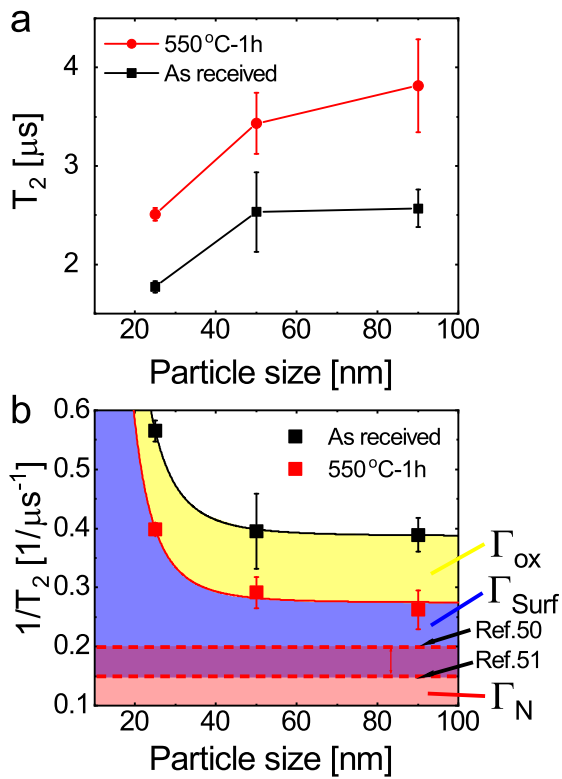}
 \caption{Particle size dependence of ${\rm T_2}$ coherence time of NV centers. 
(a) The mean ${\rm T_2}$ of different sized NDs of as-received (black) and aerobic oxidation at 550 $^\circ$C for 1 h (red). 
The error bars are 1$\sigma$ of 20 particles for 25 nm, and 5 particles for 50 and 90 nm. 
(b) The inverse of ${\rm T_2}$ (decoherence rate) of different ND-sized particles for the as-received and aerobic oxidation at 550 $^\circ$C for 1 h. 
$\Gamma_{\rm N}$, $\Gamma_{\rm Ox}$, and $\Gamma_{\rm Surf}$ are the decoherence contributions from nearby P$_1$ centers, the sources removed by aerobic oxidation, and other sources intrinsic to NDs, respectively.}
 \label{fgr:figure4}
\end{figure}

To obtain further insight into the effect of surface oxidation on ${\rm T_2}$, we measure the ND-size dependence of ${\rm T_2}$ for as-received and aerobic oxidation at 550 $\si{\degreeCelsius}$ for 1 h. 
Specifically, the ND sizes of 50 and 90 nm samples are aerobically oxidized at 550 $\si{\degreeCelsius}$ for an hour and measured to obtain the mean ${\rm T_2}$ in comparison with the as-received ND samples. 
Figure~\ref{fgr:figure4}a shows a dependence of ${\rm T_2}$ on the ND size for the two different surface NDs. 
Interestingly, ${\rm T_2}$ is increased more significantly in the larger ND samples.
While surface oxidation would have more impact on smaller NDs owing to the small distance between the surface and NV centers, the aerobic oxidation seems to more significantly improve ${\rm T_2}$ in the larger ND samples. 

This tendency can be understood by analyzing their decoherence sources. 
The observed decoherence rates are the sum of all possible decoherence contributions, which we describe here as 
\begin{equation}
    \Gamma_2 \\ (\\ = \frac{1}{\rm T_2})=\Gamma_{\rm N}+\Gamma_{\rm Ox}+\Gamma_{\rm Surf}, 
    \label{eq:1}
\end{equation}
where $\Gamma_{\rm N}$, $\Gamma_{\rm Ox}$, and $\Gamma_{\rm Surf}$ are the decoherence contributions from nearby  N-impurity centers, the sources removed by the aerobic oxidation, and other decoherence sources on the surface, respectively. 

Figure~\ref{fgr:figure4}b shows a plot of the decoherence rates (${\rm T_2^{-1}}$) as a function of ND size with visualization of the amount of the above three decoherence sources.
${\rm \Gamma_{N}}$ only depends on the nitrogen concentration of the diamonds (red shaded area) and not on the size. 
We set ${\rm \Gamma_{N}}$ = 0.15 $\si{\us}^{-1}$ or 0.2 $\si{\us}^{-1}$ by referring to the previous reports using bulk diamonds with similar nitrogen impurity concentrations (see Table~S2 for the summary of the previously reported ${\rm T_2}$).~\cite{rondin2014magnetometry,pham2013magnetic,takahashi2008quenching} 
${\rm \Gamma_{Ox}}$ is the one we have removed by surface oxidation (yellow area), which results in a difference of ${\rm T_2}$ between the as-received and surface oxidized samples. 
The remaining contribution is $\Gamma_{\rm Surf}$ that are all other decoherence sources created during the morphology transitions from bulk to NDs.

%0.2の方はPham2013.0,15の方はtakahashi2008

Importantly, we find that $\Gamma_{\rm Ox}$ does not depend on ND size and has a significant influence even on the 200-nm ND samples. 
In case of bulk diamonds, the spin properties are not affected as long as the NV center is deeper than 70 nm~\cite{ishikawa2012optical}, and only very shallow NV centers (for example 5 nm) show strong dependency on surface termination.~\cite{ohashi2013negatively}
Assuming ND morphology as a sphere and NV distribution to be random in the NDs, the NV-surface distance is half of the ND size at the longest.
Surface oxidation, therefore, is not expected to influence the large ND sample of 90 nm, significantly.
The present observation of the constant influence of $\Gamma_{\rm Ox}$ irrespective of the ND size is a new important knowledge toward realizing an extended ${\rm T_2}$ in NDs. 

%Note that $\Gamma_{\rm Ox}$ of the smallest ND size of 25 nm may be more influenced than the larger ND samples because some NV centers must be located at a distance of less than 10 nm from the surface. 
%The present data are, however, not sufficient to correctly indicate the dependence of $\Gamma_{\rm Ox}$ in a range of less than 10 nm of the NV-surface distance owing to the very small probability of finding NV centers in such small ND samples.

A second important finding of the present experiment is that $\Gamma_{\rm Surf}$ is the one showing a strong size-dependence of ${\rm T_2}$ of NV centers in NDs.
This size-dependence can be understood by considering the interactions between the NV spin and surface spins that provide $d_0^{-4}$ dependence~\cite{tetienne2013spin,myers2014probing,song2014statistical}, where $d_0$ is the diameter of the NDs. 
The $d_0^{-4}$ dependence shows a significant increase only in the range $d_0 < 40$ nm, which is consistent with the present observation, where the 25-nm-sized NDs show 1.3 times increase in $\Gamma_{\rm Surf}$ from the 50-nm NDs, whereas the 50-nm-sized NDs show almost the same $\Gamma _{\rm surf}$ as that of the 90-nm-sized NDs within the error range.
The fitting of $d_0^{-4}$ to the data in Fig.~\ref{fgr:figure4}b indeed shows a good agreement.%
The remaining offset contribution of $\Gamma _{\rm surf}$ come from other surface decoherence sources, such as the carboxyl terminate recently found as a decoherence source~\cite{ryan2018impact}. 
Further detailed analysis on the surface decoherence sources can help the ND-NV ${\rm T_2}$ to reach the ${\rm \Gamma_{N}}$ limit, which may be the previous case for Type-IIa CVD-grown NDs with a size of 50--150 nm, for example (see Table S2).~\cite{trusheim2013scalable}

%The present experimental demonstration indicates that the surface effect hitherto claimed to strongly decohere the NV spin systems in the NDs can only be important in small NDs with a size smaller than $\sim$ 50 nm.
%In contrast, ${\rm T_2}$ of NV centers in the large NDs ($>$ 50 nm) can be significantly prolonged that may reach the ${\rm \Gamma_{N}}$ limit by the aerobic oxidation, 
%which may be the previous case for Type-IIa CVD-grown NDs with a size of 50--150 nm, for example (see Table S2).~\cite{trusheim2013scalable}
These findings are particularly important for quantum nanoscale sensing when choosing an appropriate ND size for sensing applications.
For example, NDs used for quantum thermometry do not require a very small size like single digit nanometers because the spatial resolution of temperature is mainly limited by diffusion, and hence, the ND size of 50--100 nm may be better for the thermometry applications to maximally harness the long ${\rm T_2}$.
This size range also fits the basic mechanism of cellular uptake of nanoparticles, such as clathrin-mediated endocytosis.
~\cite{conner2003regulated}
In contrast, the magnetometry application requires a very small ND size with long ${\rm T_2}$ to obtain the high sensitivity of magnetic sensing, which is more challenging to achieve without understanding the origin of $\Gamma_{\rm Surf}$. 

Note that determining ${\rm \Gamma_{N}}$ of the present ND samples is a great challenge and is yet to be achieved. 
The original bulk material of the as-received NDs has a size of $\sim$ 300 $\si{\um}$ (MSY60/80, Microdiamant), and we attempt to measure ${\rm T_2}$ of these micro-crystals (see Fig.~S6).
However, it was impossible to measure any pulsed ODMR signal most probably because of very short spin coherence time owing to the strong interactions between nearby spin decoherence sources such as nearby NV centers and N-impurity centers. 
In fact, the CW-ODMR spectra of these microcrystals are significantly broadened, while the NV centers in their single nanoparticles can allow pulsed ODMR signals because other noise sources vanish owing to the down-sizing.
The reported ${\rm T_2}$ values of ensemble NV centers in Type-Ib bulk diamonds (with similar nitrogen impurity of $\sim$ 100 ppm) in the literature show a broad range of 1--7 $\si{\us}$.
Among them, we take 6.7 $\si{\us}$ and 5 $\si{\us}$ as the most probable limit for the measurement data presented in Fig.~\ref{fgr:figure4}b. 

At present, we cannot experimentally clarify the sources of $\Gamma_{\rm Surf}$.
A possible approach to investigate the mechanism is using a noise spectroscopy of the surface-induced decoherence of NV centers in NDs that can characterize the noise species on the diamond surface, as demonstrated for shallow NV centers near the bulk diamond surface.~\cite{ohno2012engineering,kim2014effect} 
Applying hydrogen termination to the surface-oxidized ND surface can become a starting point of such noise spectroscopy, because the hydrogen termination gives a characteristic and simple FTIR spectra, thereby defining the surface properties.~\cite{takimoto2010preparation}
The hydrogen-terminated surface can be then converted to oxygen termination (by either aerobic or acidic oxidation). 
The complicated signatures of the surface oxidation in FTIR spectra owing to the various functional groups (ester, carbonyl, carboxyl, and hydroxide) can be readily understood in comparison with hydrogen termination.
Such detailed study on the relation between NV decoherence and surface properties is necessary for the further extension of ${\rm T_2}$ in small NDs.

\section{Conclusions}
In conclusion, we studied the effects of ND surface properties to extend ${\rm T_2}$ of single NV centers in Type-Ib NDs by systematically analyzing the relation between ${\rm T_2}$ and the surface properties for aerobic and acidic oxidations.  
By employing the statistical approach, we found that aerobic oxidation most effectively oxidized the surface and extended ${\rm T_2}$ by a factor of 1.6 to the original NDs. 
The ND-size dependence of ${\rm T_2}$ clearly showed that surface oxidation removes a constant decoherence contribution irrespective of the ND size in the present ND size range and that the new surface-derived decoherence source emerges when the ND size reduces below 50 nm. 
The present results provide quantitative information on the decoherence sources of the NV spin systems in NDs and will enable a strategic surface modification toward better spin manipulations of NV centers in the context of quantum information science and nanoscale quantum sensing.

\section{Methods}
\subsubsection{Sample preparation}
We used commercially available NDs (Microdiamant, MSY 0-0.05, Type Ib, HPHT) as a starting material. These NDs are a complex mixtures of diamond and graphitic carbons, and thus, are black.
The NDs were centrifuged, and the supernatant was replaced with pure water to reduce sp$^2$-like materials that showed background fluorescence.
We repeated this washing process five times to purify the NDs.
The purified NDs were sonicated for an hour to obtain colloidal dispersions (as-received).
%Since the background was too strong when used as it was, the concentration was adjusted (ND 100 $\mu$L: 200 $\mu$L of water).
%
We oxidized the ND surface by aerobic annealing or solution-based acid cleaning. For aerobic annealing, the as-received NDs were drop-casted on glass substrates and completely dried. 
They were heated in an oven at 550 $\si{\degreeCelsius}$ under the atmospheric pressure of air. 
The annealed ND powders were re-dispersed in distilled water.
For acid-cleaning, the as-received NDs were processed in a mixture of sulfuric (98 \%), nitric (97 \%), and perchloric (60 \%) acids (volume ratio 1:1:1) at 90 $\si{\degreeCelsius}$, followed by a base wash with 0.1-M NaOH and 0.1-M HCl. 
The NDs were rinsed five times with distilled water. 
The resultant ND dispersion turned white, indicating the removal of the graphite layers.
%\textcolor{red}{ what is this meaning?? Since the background was too strong when used as it was, the concentration was adjusted (ND 100 $\mu$L: 200 $\mu$L of water), a sample was obtained in the same approach as “as-received”.}
%%%%%%%%%%%%%%%%%%%%%%%%%%%%%%%%%%%%%%%%%%%%%%%%%%%%%%%%%%%%%%%%%%%%%%%%%%%%%%%%%%%%%%%%%%%
%\textcolor{red}{ what is this meaning?? Since the background was too strong when used as it was, the concentration was adjusted (ND 100 $\mu$L: 200 $\mu$L of water), a sample was obtained in the same approach as “as-received”.}
%実験ノートに、そのまま使うと蛍光バックグラウンドが強いから、水と薄める。と書いてあったので記載しました。
%実験をし始めた初期の頃だと思います。
%as-receivedでも薄めているので、同様の方法を使用したとなっているのかと思います。（おそらくですが、元々はas-receivedのSample preparationの項にも書いていたのだと思います。）
%
%washing procedure を書いて下さい、お願いします
%%%%%%%%%%%%%%%%%%%%%%%%%%%%%%%%%%%%%%%%%%%%%%%%%%%%%%%%%%%%%%%%%%%%%%%%%%%%%%%%%%%%%%%%%%%
%
Polyvinyl alcohol was added to ND dispersion and sonicated for 5 min to create a uniform film in the spin-coating process.
A small droplet of the dispersion was spin-coated to cleaned coverslips.
The topographies of the spin-coated samples were obtained using an AFM (Bruker AFM).
These data were used to obtain the particle height distributions by taking the particle height as the particle size because only the vertical information provides the correct particle size owing to the effect of the cantilever size.

\subsubsection{Surface characterizations}
We analyzed the ND surface through Raman spectroscopy, (XPS), and (FTIR). 
The ND dispersions were drop-casted and dried on silicon wafers for the Raman and XPS measurements. 
Raman spectra were measured using the excitation wavelength at 325 nm (HORIBA, LabRAM HR Evolution and Renishaw, InVia Reflex Spectrometer System).
XPS was performed with an AlK$\alpha$ spectral line of the XPS spectrometer (JEOL, JPS-9200). 
Before the XPS measurements, electron gun neutralization was applied to discharge the samples.
For the FTIR measurements, the dried powder samples were placed on top of the ZnSe prism of the FTIR-ATR spectrometer (JUSCO, FTIR, ATR) and measured.

\subsubsection{ODMR}
We used a home-built confocal microscope system that integrates a microwave circuit for the spin control (Fig.~S3). 
A continuous-wave 532-nm laser was used for the excitation with a typical excitation intensity of $\sim$ 90 kW/cm$^2$ for acquiring the scanning image and second-order photon correlation histograms, which was near the fluorescence saturation laser intensity. 
An oil-immersion objective (numerical aperture: 1.4) was used for both the excitation and fluorescence collection. 
NV fluorescence was filtered by a dichroic beam splitter and a long pass filter. 
The fluorescence was then coupled to an optical fiber having a core diameter of $\sim$ 10 $\si{\um}$. 
The fiber-coupled fluorescence was guided into a Hanbury-Brown-Twiss (HBT) setup, which consists of two avalanche photodiodes and a 50:50 beam splitter.
For the spectral measurements, the microscope was connected to a fiber-coupled spectrometer equipped with a liquid-nitrogen-cooled charge-coupled device (CCD) camera. 
A time-correlated single-photon counting module (PicoQuant, TimeHarp-260) was used to obtain second-order photon correlation histograms to identify single NV centers by measuring the antibunching.
Microwaves were generated from a source (Rohde\&Schwarz, SMB100A) and amplified by 45 dB (Mini-circuit, ZHL-16W-43+). 
They were then fed to a microwave linear antenna placed on the coverslips.
The typical microwave excitation power for the continuous-wave ODMR spectral measurement was 35 dBm (3.2 W) under the impedance mismatching condition. 
The avalanche photodiode (APD) detection was gated by using a radio-frequency (RF) switch for the microwave ON and OFF triggered by a bit pattern generator (PBESR-PRO-300, Spincore).
For the ODMR measurements, the optical excitation intensity was reduced to the range of 10--40 kW/cm$^2$ to avoid the optical decoherence of the NV spins.

A small magnetic field ($\sim$30 Gauss) was applied to split the magnetic sublevels to the extent that the peaks are well-separated in frequency. 
We employed Rabi and spin-echo sequences in the pulsed ODMR measurements.
The spin-echo sequence determines the spin coherence time (${\rm T_2}$). 
We measured both $\pi$/2--$\pi$--$\pi$/2 and $\pi$/2--$\pi$--3$\pi$/2 sequences and subtracted these signals from each other to cancel the common mode noise in the spin echo measurements.~\cite{bolshedvorskii2017single}

In the measurements of the ND size dependence of $T_{\rm 2}$, most of the NV centers found in the 50-nm-sized NDs were single and proceeded to the subsequent ODMR measurements (see Fig.~S5 for the scanning image).
The 90-nm-sized NDs contained several large ND particles (or agglomeration) and showed bright fluorescence.
These ND particles did not show ODMR signals. 
We, therefore, searched relatively dark fluorescence spots, which show photon counts of single or double NV centers, and measured the ODMR signals.

\begin{acknowledgement}
We thank Prof. Noboru Ohtani for the AFM measurements, 
Mr. Keisuke Oshimi for the computational support of ${\rm T_2}$ measurements,  
and HORIBA Inc. for the part of the Raman measurements.
We also thank Dr. Akihiro Shimizu for the help with XPS measurements.
A part of this study was conducted at Hokkaido University and Chitose Institute of Science and Technology, supported by ``Nanotechnology Platform'' Program of the Ministry of Education, Culture, Sports, Science and Technology (MEXT), Japan.
MF acknowledges the financial support provided by JSPS-KAKENHI (Nos. 26706007, 26610077, 16K13646, and 17H02741), MEXT-LEADER program, and Osaka City University (OCU-Strategic Research Grant 2017 \& 2018 for young researchers and top-priority research).
SS acknowledges the financial support provided by JSPS-KAKENHI (No. 26220602).
HH thanks JSPS KAKENHI, Grant-in-Aids for Basic Research (B) (No. 16H04181) and Scientific Research on Innovative Areas ``Innovations for Light-Energy Conversion (I$^4$LEC)'' (Nos. 17H06433, 17H0637) for the financial support.

\end{acknowledgement}

\section{Author Information}
\subsection{Corresponding Authors}
*E-mail: masazumi@osaka-cu.ac.jp \\
*E-mail: SShikata@kwansei.ac.jp

\subsection{Author Contributions}
MF, SS designed the research.
RT, MF, YN conducted the optical experiments.
RT, MF, YSR, YN, YSG conducted the sample preparation and material characterization.
RT, MF, SS, HH wrote the manuscript.
All authors contributed to the discussion. 

\subsection{Notes}
The authors declare no competing interest.

%\subsection{Emails of all the authors}
%Masazumi Fujiwara: masazumi@sci.osaka-cu.ac.jp \\
%Oliver Neitzke: oliver.neitzke@physik.hu-berlin.de \\
%Tim Schr{\"o}der: schroder@nbi.ku.dk \\
%Andreas W. Schell: andreas.schell@icfo.eu \\
%Janik Wolters: janik.wolters@physik.hu-berlin.de \\
%Jiabao Zheng: jz2466@columbia.edu \\
%Sara Mouradian: smouradi@mit.edu \\
%Mohamed Almoktar: malmokhtar00@gmail.com \\
%Shigeki Takeuchi: takeuchi@kuee.kyoto-u.ac.jp\\
%Dirk Englund: englund@mit.edu \\
%Oliver Benson: oliver.benson@physik.hu-berlin.de\\

\begin{suppinfo}
The Supporting Information is available free of charge on the ACS Publications website.
Particle size determination using AFM, statistical figures of particle size distributions, peak assignment of FTIR spectra, experimental setup for ODMR measurements, typical dataset of the confocal microscope and ODMR measurements, summary of the reported spin coherence times of various diamonds, brightness of NV centers in the ND samples of 50 and 90 nm size, and ODMR spectra of NV centers in micron-sized crystals (PDF).
%The Supporting Information is available free of charge on the ACS Publications website.
%More information on the ND size measurement, protocol, ODMR measurement presented in this work (PDF). 
\end{suppinfo}

\begin{tocentry}
\includegraphics{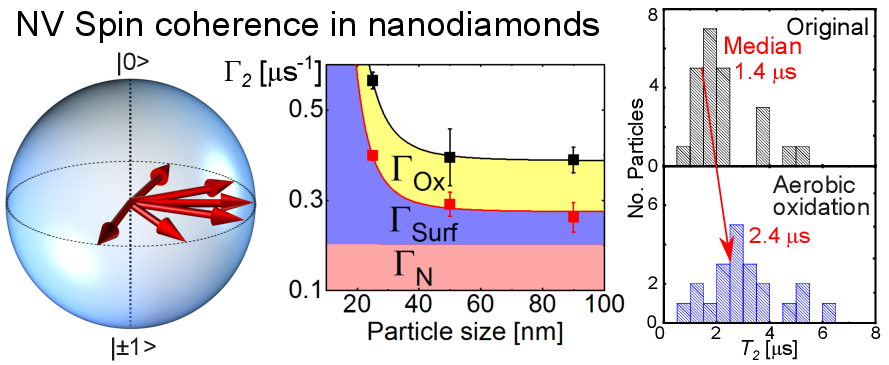}
Some text to explain the graphic.
\end{tocentry}

%\bibliography{achemso-demo}

\begin{mcitethebibliography}{60}
\providecommand*\natexlab[1]{#1}
\providecommand*\mciteSetBstSublistMode[1]{}
\providecommand*\mciteSetBstMaxWidthForm[2]{}
\providecommand*\mciteBstWouldAddEndPuncttrue
  {\def\EndOfBibitem{\unskip.}}
\providecommand*\mciteBstWouldAddEndPunctfalse
  {\let\EndOfBibitem\relax}
\providecommand*\mciteSetBstMidEndSepPunct[3]{}
\providecommand*\mciteSetBstSublistLabelBeginEnd[3]{}
\providecommand*\EndOfBibitem{}
\mciteSetBstSublistMode{f}
\mciteSetBstMaxWidthForm{subitem}{(\alph{mcitesubitemcount})}
\mciteSetBstSublistLabelBeginEnd
  {\mcitemaxwidthsubitemform\space}
  {\relax}
  {\relax}

\bibitem[Togan \latin{et~al.}(2010)Togan, Chu, Trifonov, Jiang, Maze,
  Childress, Dutt, S{\o}rensen, Hemmer, Zibrov, \latin{et~al.}
  others]{togan2010quantum}
Togan,~E.; Chu,~Y.; Trifonov,~A.; Jiang,~L.; Maze,~J.; Childress,~L.;
  Dutt,~M.~G.; S{\o}rensen,~A.~S.; Hemmer,~P.; Zibrov,~A.~S. \latin{et~al.}
  Quantum entanglement between an optical photon and a solid-state spin qubit.
  \emph{Nature} \textbf{2010}, \emph{466}, 730\relax
\mciteBstWouldAddEndPuncttrue
\mciteSetBstMidEndSepPunct{\mcitedefaultmidpunct}
{\mcitedefaultendpunct}{\mcitedefaultseppunct}\relax
\EndOfBibitem
\bibitem[Kosaka and Niikura(2015)Kosaka, and Niikura]{kosaka2015entangled}
Kosaka,~H.; Niikura,~N. Entangled absorption of a single photon with a single
  spin in diamond. \emph{Physical Review Letters} \textbf{2015}, \emph{114},
  053603\relax
\mciteBstWouldAddEndPuncttrue
\mciteSetBstMidEndSepPunct{\mcitedefaultmidpunct}
{\mcitedefaultendpunct}{\mcitedefaultseppunct}\relax
\EndOfBibitem
\bibitem[Almokhtar \latin{et~al.}(2014)Almokhtar, Fujiwara, Takashima, and
  Takeuchi]{almokhtar2014numerical}
Almokhtar,~M.; Fujiwara,~M.; Takashima,~H.; Takeuchi,~S. Numerical simulations
  of nanodiamond nitrogen-vacancy centers coupled with tapered optical fibers
  as hybrid quantum nanophotonic devices. \emph{Optics Express} \textbf{2014},
  \emph{22}, 20045--20059\relax
\mciteBstWouldAddEndPuncttrue
\mciteSetBstMidEndSepPunct{\mcitedefaultmidpunct}
{\mcitedefaultendpunct}{\mcitedefaultseppunct}\relax
\EndOfBibitem
\bibitem[Fujiwara \latin{et~al.}(2015)Fujiwara, Zhao, Noda, Ikeda, Sumiya, and
  Takeuchi]{fujiwara2015ultrathin}
Fujiwara,~M.; Zhao,~H.-Q.; Noda,~T.; Ikeda,~K.; Sumiya,~H.; Takeuchi,~S.
  Ultrathin fiber-taper coupling with nitrogen vacancy centers in nanodiamonds
  at cryogenic temperatures. \emph{Optics Letters} \textbf{2015}, \emph{40},
  5702--5705\relax
\mciteBstWouldAddEndPuncttrue
\mciteSetBstMidEndSepPunct{\mcitedefaultmidpunct}
{\mcitedefaultendpunct}{\mcitedefaultseppunct}\relax
\EndOfBibitem
\bibitem[Schr{\"o}der \latin{et~al.}(2012)Schr{\"o}der, Fujiwara, Noda, Zhao,
  Benson, and Takeuchi]{schroder2012nanodiamond}
Schr{\"o}der,~T.; Fujiwara,~M.; Noda,~T.; Zhao,~H.-Q.; Benson,~O.; Takeuchi,~S.
  A nanodiamond-tapered fiber system with high single-mode coupling efficiency.
  \emph{Optics Express} \textbf{2012}, \emph{20}, 10490--10497\relax
\mciteBstWouldAddEndPuncttrue
\mciteSetBstMidEndSepPunct{\mcitedefaultmidpunct}
{\mcitedefaultendpunct}{\mcitedefaultseppunct}\relax
\EndOfBibitem
\bibitem[Fujiwara \latin{et~al.}(2016)Fujiwara, Yoshida, Noda, Takashima,
  Schell, Mizuochi, and Takeuchi]{fujiwara2016manipulation}
Fujiwara,~M.; Yoshida,~K.; Noda,~T.; Takashima,~H.; Schell,~A.~W.;
  Mizuochi,~N.; Takeuchi,~S. Manipulation of single nanodiamonds to ultrathin
  fiber-taper nanofibers and control of NV-spin states toward fiber-integrated
  $\lambda$-systems. \emph{Nanotechnology} \textbf{2016}, \emph{27},
  455202\relax
\mciteBstWouldAddEndPuncttrue
\mciteSetBstMidEndSepPunct{\mcitedefaultmidpunct}
{\mcitedefaultendpunct}{\mcitedefaultseppunct}\relax
\EndOfBibitem
\bibitem[Doherty \latin{et~al.}(2013)Doherty, Manson, Delaney, Jelezko,
  Wrachtrup, and Hollenberg]{doherty2013nitrogen}
Doherty,~M.~W.; Manson,~N.~B.; Delaney,~P.; Jelezko,~F.; Wrachtrup,~J.;
  Hollenberg,~L.~C. The nitrogen-vacancy colour centre in diamond.
  \emph{Physics Reports} \textbf{2013}, \emph{528}, 1--45\relax
\mciteBstWouldAddEndPuncttrue
\mciteSetBstMidEndSepPunct{\mcitedefaultmidpunct}
{\mcitedefaultendpunct}{\mcitedefaultseppunct}\relax
\EndOfBibitem
\bibitem[Schirhagl \latin{et~al.}(2014)Schirhagl, Chang, Loretz, and
  Degen]{schirhagl2014nitrogen}
Schirhagl,~R.; Chang,~K.; Loretz,~M.; Degen,~C.~L. Nitrogen-vacancy centers in
  diamond: nanoscale sensors for physics and biology. \emph{Annual Review of
  Physical Chemistry} \textbf{2014}, \emph{65}, 83--105\relax
\mciteBstWouldAddEndPuncttrue
\mciteSetBstMidEndSepPunct{\mcitedefaultmidpunct}
{\mcitedefaultendpunct}{\mcitedefaultseppunct}\relax
\EndOfBibitem
\bibitem[Iwasaki \latin{et~al.}(2017)Iwasaki, Naruki, Tahara, Makino, Kato,
  Ogura, Takeuchi, Yamasaki, and Hatano]{iwasaki2017direct}
Iwasaki,~T.; Naruki,~W.; Tahara,~K.; Makino,~T.; Kato,~H.; Ogura,~M.;
  Takeuchi,~D.; Yamasaki,~S.; Hatano,~M. Direct nanoscale sensing of the
  internal electric field in operating semiconductor devices using single
  electron spins. \emph{ACS Nano} \textbf{2017}, \emph{11}, 1238--1245\relax
\mciteBstWouldAddEndPuncttrue
\mciteSetBstMidEndSepPunct{\mcitedefaultmidpunct}
{\mcitedefaultendpunct}{\mcitedefaultseppunct}\relax
\EndOfBibitem
\bibitem[Neumann \latin{et~al.}(2013)Neumann, Jakobi, Dolde, Burk, Reuter,
  Waldherr, Honert, Wolf, Brunner, Shim, Suter, Sumiya, Isoya, and
  Wrachtrup]{neumann2013high}
Neumann,~P.; Jakobi,~I.; Dolde,~F.; Burk,~C.; Reuter,~R.; Waldherr,~G.;
  Honert,~J.; Wolf,~T.; Brunner,~A.; Shim,~J.~H. \latin{et~al.}  High-precision
  nanoscale temperature sensing using single defects in diamond. \emph{Nano
  Letters} \textbf{2013}, \emph{13}, 2738--2742\relax
\mciteBstWouldAddEndPuncttrue
\mciteSetBstMidEndSepPunct{\mcitedefaultmidpunct}
{\mcitedefaultendpunct}{\mcitedefaultseppunct}\relax
\EndOfBibitem
\bibitem[Andrich \latin{et~al.}(2017)Andrich, Charles, Liu, Bretscher, Berman,
  Heremans, Nealey, and Awschalom]{andrich2017long}
Andrich,~P.; Charles,~F.; Liu,~X.; Bretscher,~H.~L.; Berman,~J.~R.;
  Heremans,~F.~J.; Nealey,~P.~F.; Awschalom,~D.~D. Long-range spin wave
  mediated control of defect qubits in nanodiamonds. \emph{npj Quantum
  Information} \textbf{2017}, \emph{3}, 28\relax
\mciteBstWouldAddEndPuncttrue
\mciteSetBstMidEndSepPunct{\mcitedefaultmidpunct}
{\mcitedefaultendpunct}{\mcitedefaultseppunct}\relax
\EndOfBibitem
\bibitem[Bose \latin{et~al.}(2017)Bose, Mazumdar, Morley, Ulbricht,
  Toro\ifmmode~\check{s}\else \v{s}\fi{}, Paternostro, Geraci, Barker, Kim, and
  Milburn]{PhysRevLett.119.240401}
Bose,~S.; Mazumdar,~A.; Morley,~G.~W.; Ulbricht,~H.;
  Toro\ifmmode~\check{s}\else \v{s}\fi{},~M.; Paternostro,~M.; Geraci,~A.~A.;
  Barker,~P.~F.; Kim,~M.~S.; Milburn,~G. Spin Entanglement Witness for Quantum
  Gravity. \emph{Phys. Rev. Lett.} \textbf{2017}, \emph{119}, 240401\relax
\mciteBstWouldAddEndPuncttrue
\mciteSetBstMidEndSepPunct{\mcitedefaultmidpunct}
{\mcitedefaultendpunct}{\mcitedefaultseppunct}\relax
\EndOfBibitem
\bibitem[Rabeau \latin{et~al.}(2007)Rabeau, Stacey, Rabeau, Prawer, Jelezko,
  Mirza, and Wrachtrup]{rabeau2007single}
Rabeau,~J.; Stacey,~A.; Rabeau,~A.; Prawer,~S.; Jelezko,~F.; Mirza,~I.;
  Wrachtrup,~J. Single nitrogen vacancy centers in chemical vapor deposited
  diamond nanocrystals. \emph{Nano Letters} \textbf{2007}, \emph{7},
  3433--3437\relax
\mciteBstWouldAddEndPuncttrue
\mciteSetBstMidEndSepPunct{\mcitedefaultmidpunct}
{\mcitedefaultendpunct}{\mcitedefaultseppunct}\relax
\EndOfBibitem
\bibitem[Wang \latin{et~al.}(2016)Wang, Zhang, Zhang, You, Li, Guo, Feng, Song,
  Lou, Zhu, and Wang]{wang2016coherence}
Wang,~J.; Zhang,~W.; Zhang,~J.; You,~J.; Li,~Y.; Guo,~G.; Feng,~F.; Song,~X.;
  Lou,~L.; Zhu,~W. \latin{et~al.}  Coherence times of precise depth controlled
  NV centers in diamond. \emph{Nanoscale} \textbf{2016}, \emph{8},
  5780--5785\relax
\mciteBstWouldAddEndPuncttrue
\mciteSetBstMidEndSepPunct{\mcitedefaultmidpunct}
{\mcitedefaultendpunct}{\mcitedefaultseppunct}\relax
\EndOfBibitem
\bibitem[Knowles \latin{et~al.}(2014)Knowles, Kara, and
  Atat{\"u}re]{knowles2014observing}
Knowles,~H.~S.; Kara,~D.~M.; Atat{\"u}re,~M. Observing bulk diamond spin
  coherence in high-purity nanodiamonds. \emph{Nature Materials} \textbf{2014},
  \emph{13}, 21\relax
\mciteBstWouldAddEndPuncttrue
\mciteSetBstMidEndSepPunct{\mcitedefaultmidpunct}
{\mcitedefaultendpunct}{\mcitedefaultseppunct}\relax
\EndOfBibitem
\bibitem[Tisler \latin{et~al.}(2009)Tisler, Balasubramanian, Naydenov, Kolesov,
  Grotz, Reuter, Boudou, Curmi, Sennour, Thorel, B\"{o}rsch, Aulenbacher,
  Erdmann, Hemmer, Jelezko, and Wrachtrup]{tisler2009fluorescence}
Tisler,~J.; Balasubramanian,~G.; Naydenov,~B.; Kolesov,~R.; Grotz,~B.;
  Reuter,~R.; Boudou,~J.-P.; Curmi,~P.~A.; Sennour,~M.; Thorel,~A.
  \latin{et~al.}  Fluorescence and spin properties of defects in single digit
  nanodiamonds. \emph{ACS Nano} \textbf{2009}, \emph{3}, 1959--1965\relax
\mciteBstWouldAddEndPuncttrue
\mciteSetBstMidEndSepPunct{\mcitedefaultmidpunct}
{\mcitedefaultendpunct}{\mcitedefaultseppunct}\relax
\EndOfBibitem
\bibitem[Ohashi \latin{et~al.}(2013)Ohashi, Rosskopf, Watanabe, Loretz, Tao,
  Hauert, Tomizawa, Ishikawa, Ishi-Hayase, Shikata, Degen, and
  Itoh]{ohashi2013negatively}
Ohashi,~K.; Rosskopf,~T.; Watanabe,~H.; Loretz,~M.; Tao,~Y.; Hauert,~R.;
  Tomizawa,~S.; Ishikawa,~T.; Ishi-Hayase,~J.; Shikata,~S. \latin{et~al.}
  Negatively charged nitrogen-vacancy centers in a 5 nm thin 12C diamond film.
  \emph{Nano Letters} \textbf{2013}, \emph{13}, 4733--4738\relax
\mciteBstWouldAddEndPuncttrue
\mciteSetBstMidEndSepPunct{\mcitedefaultmidpunct}
{\mcitedefaultendpunct}{\mcitedefaultseppunct}\relax
\EndOfBibitem
\bibitem[Trusheim \latin{et~al.}(2013)Trusheim, Li, Laraoui, Chen, Bakhru,
  Schröder, Gaathon, Meriles, and Englund]{trusheim2013scalable}
Trusheim,~M.~E.; Li,~L.; Laraoui,~A.; Chen,~E.~H.; Bakhru,~H.; Schröder,~T.;
  Gaathon,~O.; Meriles,~C.~A.; Englund,~D. Scalable fabrication of high purity
  diamond nanocrystals with long-spin-coherence nitrogen vacancy centers.
  \emph{Nano Letters} \textbf{2013}, \emph{14}, 32--36\relax
\mciteBstWouldAddEndPuncttrue
\mciteSetBstMidEndSepPunct{\mcitedefaultmidpunct}
{\mcitedefaultendpunct}{\mcitedefaultseppunct}\relax
\EndOfBibitem
\bibitem[Song \latin{et~al.}(2014)Song, Zhang, Feng, Wang, Zhang, Lou, Zhu, and
  Wang]{song2014statistical}
Song,~X.; Zhang,~J.; Feng,~F.; Wang,~J.; Zhang,~W.; Lou,~L.; Zhu,~W.; Wang,~G.
  A statistical correlation investigation for the role of surface spins to the
  spin relaxation of nitrogen vacancy centers. \emph{AIP Advances}
  \textbf{2014}, \emph{4}, 047103\relax
\mciteBstWouldAddEndPuncttrue
\mciteSetBstMidEndSepPunct{\mcitedefaultmidpunct}
{\mcitedefaultendpunct}{\mcitedefaultseppunct}\relax
\EndOfBibitem
\bibitem[Laraoui \latin{et~al.}(2012)Laraoui, Hodges, and
  Meriles]{laraoui2012nitrogen}
Laraoui,~A.; Hodges,~J.~S.; Meriles,~C.~A. Nitrogen-vacancy-assisted
  magnetometry of paramagnetic centers in an individual diamond nanocrystal.
  \emph{Nano Letters} \textbf{2012}, \emph{12}, 3477--3482\relax
\mciteBstWouldAddEndPuncttrue
\mciteSetBstMidEndSepPunct{\mcitedefaultmidpunct}
{\mcitedefaultendpunct}{\mcitedefaultseppunct}\relax
\EndOfBibitem
\bibitem[Boudou \latin{et~al.}(2013)Boudou, Tisler, Reuter, Thorel, Curmi,
  Jelezko, and Wrachtrup]{boudou2013fluorescent}
Boudou,~J.-P.; Tisler,~J.; Reuter,~R.; Thorel,~A.; Curmi,~P.~A.; Jelezko,~F.;
  Wrachtrup,~J. Fluorescent nanodiamonds derived from HPHT with a size of less
  than 10 nm. \emph{Diamond and Related Materials} \textbf{2013}, \emph{37},
  80--86\relax
\mciteBstWouldAddEndPuncttrue
\mciteSetBstMidEndSepPunct{\mcitedefaultmidpunct}
{\mcitedefaultendpunct}{\mcitedefaultseppunct}\relax
\EndOfBibitem
\bibitem[McGuinness \latin{et~al.}(2013)McGuinness, Hall, Stacey, Simpson,
  Hill, Cole, Ganesan, Gibson, Prawer, Mulvaney, Jelezko, Wrachtrup, Scholten,
  and Hollenberg]{mcguinness2013ambient}
McGuinness,~L.; Hall,~L.; Stacey,~A.; Simpson,~D.; Hill,~C.; Cole,~J.;
  Ganesan,~K.; Gibson,~B.; Prawer,~S.; Mulvaney,~P. \latin{et~al.}  Ambient
  nanoscale sensing with single spins using quantum decoherence. \emph{New
  Journal of Physics} \textbf{2013}, \emph{15}, 073042\relax
\mciteBstWouldAddEndPuncttrue
\mciteSetBstMidEndSepPunct{\mcitedefaultmidpunct}
{\mcitedefaultendpunct}{\mcitedefaultseppunct}\relax
\EndOfBibitem
\bibitem[Chou and Gali(2017)Chou, and Gali]{chou2017nitrogen}
Chou,~J.-P.; Gali,~A. Nitrogen-vacancy diamond sensor: novel diamond surfaces
  from ab initio simulations. \emph{MRS Communications} \textbf{2017},
  \emph{7}, 551--562\relax
\mciteBstWouldAddEndPuncttrue
\mciteSetBstMidEndSepPunct{\mcitedefaultmidpunct}
{\mcitedefaultendpunct}{\mcitedefaultseppunct}\relax
\EndOfBibitem
\bibitem[Wolcott \latin{et~al.}(2014)Wolcott, Schiros, Trusheim, Chen,
  Nordlund, Diaz, Gaathon, Englund, and Owen]{wolcott2014surface}
Wolcott,~A.; Schiros,~T.; Trusheim,~M.~E.; Chen,~E.~H.; Nordlund,~D.;
  Diaz,~R.~E.; Gaathon,~O.; Englund,~D.; Owen,~J.~S. Surface structure of
  aerobically oxidized diamond nanocrystals. \emph{The Journal of Physical
  Chemistry C} \textbf{2014}, \emph{118}, 26695--26702\relax
\mciteBstWouldAddEndPuncttrue
\mciteSetBstMidEndSepPunct{\mcitedefaultmidpunct}
{\mcitedefaultendpunct}{\mcitedefaultseppunct}\relax
\EndOfBibitem
\bibitem[Lai and Barnard(2011)Lai, and Barnard]{lai2011modeling}
Lai,~L.; Barnard,~A.~S. Modeling the thermostability of surface
  functionalisation by oxygen, hydroxyl, and water on nanodiamonds.
  \emph{Nanoscale} \textbf{2011}, \emph{3}, 2566--2575\relax
\mciteBstWouldAddEndPuncttrue
\mciteSetBstMidEndSepPunct{\mcitedefaultmidpunct}
{\mcitedefaultendpunct}{\mcitedefaultseppunct}\relax
\EndOfBibitem
\bibitem[Nguyen \latin{et~al.}(2007)Nguyen, Chang, and
  Wu]{nguyen2007adsorption}
Nguyen,~T.-T.-B.; Chang,~H.-C.; Wu,~V. W.-K. Adsorption and hydrolytic activity
  of lysozyme on diamond nanocrystallites. \emph{Diamond and Related Materials}
  \textbf{2007}, \emph{16}, 872--876\relax
\mciteBstWouldAddEndPuncttrue
\mciteSetBstMidEndSepPunct{\mcitedefaultmidpunct}
{\mcitedefaultendpunct}{\mcitedefaultseppunct}\relax
\EndOfBibitem
\bibitem[Osswald \latin{et~al.}(2006)Osswald, Yushin, Mochalin, Kucheyev, and
  Gogotsi]{osswald2006control}
Osswald,~S.; Yushin,~G.; Mochalin,~V.; Kucheyev,~S.~O.; Gogotsi,~Y. Control of
  sp2/sp3 carbon ratio and surface chemistry of nanodiamond powders by
  selective oxidation in air. \emph{Journal of the American Chemical Society}
  \textbf{2006}, \emph{128}, 11635--11642\relax
\mciteBstWouldAddEndPuncttrue
\mciteSetBstMidEndSepPunct{\mcitedefaultmidpunct}
{\mcitedefaultendpunct}{\mcitedefaultseppunct}\relax
\EndOfBibitem
\bibitem[Bradac and Osswald(2018)Bradac, and Osswald]{bradac2018effect}
Bradac,~C.; Osswald,~S. Effect of structure and composition of nanodiamond
  powders on thermal stability and oxidation kinetics. \emph{Carbon}
  \textbf{2018}, \emph{132}, 616--622\relax
\mciteBstWouldAddEndPuncttrue
\mciteSetBstMidEndSepPunct{\mcitedefaultmidpunct}
{\mcitedefaultendpunct}{\mcitedefaultseppunct}\relax
\EndOfBibitem
\bibitem[Gaebel \latin{et~al.}(2012)Gaebel, Bradac, Chen, Say, Brown, Hemmer,
  and Rabeau]{gaebel2012size}
Gaebel,~T.; Bradac,~C.; Chen,~J.; Say,~J.; Brown,~L.; Hemmer,~P.; Rabeau,~J.
  Size-reduction of nanodiamonds via air oxidation. \emph{Diamond and Related
  Materials} \textbf{2012}, \emph{21}, 28--32\relax
\mciteBstWouldAddEndPuncttrue
\mciteSetBstMidEndSepPunct{\mcitedefaultmidpunct}
{\mcitedefaultendpunct}{\mcitedefaultseppunct}\relax
\EndOfBibitem
\bibitem[Stehlik \latin{et~al.}(2015)Stehlik, Varga, Ledinsky, Jirasek,
  Artemenko, Kozak, Ondic, Skakalova, Argentero, Pennycook, Meyer, Fejfar,
  Kromka, and Rezek]{stehlik2015size}
Stehlik,~S.; Varga,~M.; Ledinsky,~M.; Jirasek,~V.; Artemenko,~A.; Kozak,~H.;
  Ondic,~L.; Skakalova,~V.; Argentero,~G.; Pennycook,~T. \latin{et~al.}  Size
  and Purity Control of HPHT Nanodiamonds down to 1 nm. \emph{The Journal of
  Physical Chemistry C} \textbf{2015}, \emph{119}, 27708--27720\relax
\mciteBstWouldAddEndPuncttrue
\mciteSetBstMidEndSepPunct{\mcitedefaultmidpunct}
{\mcitedefaultendpunct}{\mcitedefaultseppunct}\relax
\EndOfBibitem
\bibitem[Wang \latin{et~al.}(2011)Wang, Ruslinda, Ishiyama, Ishii, and
  Kawarada]{wang2011higher}
Wang,~X.; Ruslinda,~A.~R.; Ishiyama,~Y.; Ishii,~Y.; Kawarada,~H. Higher
  coverage of carboxylic acid groups on oxidized single crystal diamond (001).
  \emph{Diamond and Related Materials} \textbf{2011}, \emph{20},
  1319--1324\relax
\mciteBstWouldAddEndPuncttrue
\mciteSetBstMidEndSepPunct{\mcitedefaultmidpunct}
{\mcitedefaultendpunct}{\mcitedefaultseppunct}\relax
\EndOfBibitem
\bibitem[Fang \latin{et~al.}(2017)Fang, Zhang, Shen, Sun, Zhang, Xue, and
  Jia]{fang2017synthesis}
Fang,~C.; Zhang,~Y.; Shen,~W.; Sun,~S.; Zhang,~Z.; Xue,~L.; Jia,~X. Synthesis
  and characterization of HPHT large single-crystal diamonds under the
  simultaneous influence of oxygen and hydrogen. \emph{CrystEngComm}
  \textbf{2017}, \emph{19}, 5727--5734\relax
\mciteBstWouldAddEndPuncttrue
\mciteSetBstMidEndSepPunct{\mcitedefaultmidpunct}
{\mcitedefaultendpunct}{\mcitedefaultseppunct}\relax
\EndOfBibitem
\bibitem[Popov \latin{et~al.}(2017)Popov, Churkin, Kirichenko, Denisov,
  Ovsyannikov, Kulnitskiy, Perezhogin, Aksenenkov, and Blank]{popov2017raman}
Popov,~M.; Churkin,~V.; Kirichenko,~A.; Denisov,~V.; Ovsyannikov,~D.;
  Kulnitskiy,~B.; Perezhogin,~I.; Aksenenkov,~V.; Blank,~V. Raman spectra and
  bulk modulus of nanodiamond in a size interval of 2--5 nm. \emph{Nanoscale
  Research Letters} \textbf{2017}, \emph{12}, 561\relax
\mciteBstWouldAddEndPuncttrue
\mciteSetBstMidEndSepPunct{\mcitedefaultmidpunct}
{\mcitedefaultendpunct}{\mcitedefaultseppunct}\relax
\EndOfBibitem
\bibitem[Mochalin \latin{et~al.}(2008)Mochalin, Osswald, and
  Gogotsi]{mochalin2008contribution}
Mochalin,~V.; Osswald,~S.; Gogotsi,~Y. Contribution of functional groups to the
  Raman spectrum of nanodiamond powders. \emph{Chemistry of Materials}
  \textbf{2008}, \emph{21}, 273--279\relax
\mciteBstWouldAddEndPuncttrue
\mciteSetBstMidEndSepPunct{\mcitedefaultmidpunct}
{\mcitedefaultendpunct}{\mcitedefaultseppunct}\relax
\EndOfBibitem
\bibitem[Rondin \latin{et~al.}(2010)Rondin, Dantelle, Slablab, Grosshans,
  Treussart, Bergonzo, Perruchas, Gacoin, Chaigneau, Chang, V., and
  J-F]{rondin2010surface}
Rondin,~L.; Dantelle,~G.; Slablab,~A.; Grosshans,~F.; Treussart,~F.;
  Bergonzo,~P.; Perruchas,~S.; Gacoin,~T.; Chaigneau,~M.; Chang,~J.,~H-C
  \latin{et~al.}  Surface-induced charge state conversion of nitrogen-vacancy
  defects in nanodiamonds. \emph{Physical Review B} \textbf{2010}, \emph{82},
  115449\relax
\mciteBstWouldAddEndPuncttrue
\mciteSetBstMidEndSepPunct{\mcitedefaultmidpunct}
{\mcitedefaultendpunct}{\mcitedefaultseppunct}\relax
\EndOfBibitem
\bibitem[Laube \latin{et~al.}(2018)Laube, Hellweg, Sturm, Griebel, Grundmann,
  Kahnt, and Abel]{laube2018photo}
Laube,~C.; Hellweg,~J.; Sturm,~C.; Griebel,~J.; Grundmann,~M.; Kahnt,~A.;
  Abel,~B. Photo-Induced-Heating of Graphitized Nanodiamonds Monitored by the
  Raman-Diamond-Peak. \emph{The Journal of Physical Chemistry C} \textbf{2018},
  \relax
\mciteBstWouldAddEndPunctfalse
\mciteSetBstMidEndSepPunct{\mcitedefaultmidpunct}
{}{\mcitedefaultseppunct}\relax
\EndOfBibitem
\bibitem[Laube \latin{et~al.}(2017)Laube, Riyad, Lotnyk, Lohmann, Kranert,
  Hermann, Knolle, Oeckinghaus, Reuter, Denisenko, Kahnt, and
  Abel]{laube2017defined}
Laube,~C.; Riyad,~Y.; Lotnyk,~A.; Lohmann,~F.; Kranert,~C.; Hermann,~R.;
  Knolle,~W.; Oeckinghaus,~T.; Reuter,~R.; Denisenko,~A. \latin{et~al.}
  Defined functionality and increased luminescence of nanodiamonds for sensing
  and diagnostic applications by targeted high temperature reactions and
  electron beam irradiation. \emph{Materials Chemistry Frontiers}
  \textbf{2017}, \emph{1}, 2527--2540\relax
\mciteBstWouldAddEndPuncttrue
\mciteSetBstMidEndSepPunct{\mcitedefaultmidpunct}
{\mcitedefaultendpunct}{\mcitedefaultseppunct}\relax
\EndOfBibitem
\bibitem[Inel \latin{et~al.}(2016)Inel, Ungureau, Varley, Hirani, and
  Holt]{inel2016solvent}
Inel,~G.~A.; Ungureau,~E.-M.; Varley,~T.~S.; Hirani,~M.; Holt,~K.~B.
  Solvent--surface interactions between nanodiamond and ethanol studied with in
  situ infrared spectroscopy. \emph{Diamond and Related Materials}
  \textbf{2016}, \emph{61}, 7--13\relax
\mciteBstWouldAddEndPuncttrue
\mciteSetBstMidEndSepPunct{\mcitedefaultmidpunct}
{\mcitedefaultendpunct}{\mcitedefaultseppunct}\relax
\EndOfBibitem
\bibitem[Jiang \latin{et~al.}(1996)Jiang, Xu, and Ji]{jiang1996ftir}
Jiang,~T.; Xu,~K.; Ji,~S. FTIR studies on the spectral changes of the surface
  functional groups of ultradispersed diamond powder synthesized by explosive
  detonation after treatment in hydrogen, nitrogen, methane and air at
  different temperatures. \emph{Journal of the Chemical Society, Faraday
  Transactions} \textbf{1996}, \emph{92}, 3401--3406\relax
\mciteBstWouldAddEndPuncttrue
\mciteSetBstMidEndSepPunct{\mcitedefaultmidpunct}
{\mcitedefaultendpunct}{\mcitedefaultseppunct}\relax
\EndOfBibitem
\bibitem[Shenderova \latin{et~al.}(2011)Shenderova, Panich, Moseenkov, Hens,
  Kuznetsov, and Vieth]{shenderova2011hydroxylated}
Shenderova,~O.; Panich,~A.; Moseenkov,~S.; Hens,~S.; Kuznetsov,~V.;
  Vieth,~H.-M. Hydroxylated detonation nanodiamond: FTIR, XPS, and NMR studies.
  \emph{The Journal of Physical Chemistry C} \textbf{2011}, \emph{115},
  19005--19011\relax
\mciteBstWouldAddEndPuncttrue
\mciteSetBstMidEndSepPunct{\mcitedefaultmidpunct}
{\mcitedefaultendpunct}{\mcitedefaultseppunct}\relax
\EndOfBibitem
\bibitem[Gin{\'e}s \latin{et~al.}(2017)Gin{\'e}s, Mandal, Cheng, Sow, and
  Williams]{gines2017positive}
Gin{\'e}s,~L.; Mandal,~S.; Cheng,~C.-L.; Sow,~M.; Williams,~O.~A. Positive zeta
  potential of nanodiamonds. \emph{Nanoscale} \textbf{2017}, \emph{9},
  12549--12555\relax
\mciteBstWouldAddEndPuncttrue
\mciteSetBstMidEndSepPunct{\mcitedefaultmidpunct}
{\mcitedefaultendpunct}{\mcitedefaultseppunct}\relax
\EndOfBibitem
\bibitem[Hauf \latin{et~al.}(2011)Hauf, Grotz, Naydenov, Dankerl, Pezzagna,
  Meijer, Jelezko, Wrachtrup, Stutzmann, Reinhard, and
  Garrido]{hauf2011chemical}
Hauf,~M.; Grotz,~B.; Naydenov,~B.; Dankerl,~M.; Pezzagna,~S.; Meijer,~J.;
  Jelezko,~F.; Wrachtrup,~J.; Stutzmann,~M.; Reinhard,~F. \latin{et~al.}
  Chemical control of the charge state of nitrogen-vacancy centers in diamond.
  \emph{Physical Review B} \textbf{2011}, \emph{83}, 081304\relax
\mciteBstWouldAddEndPuncttrue
\mciteSetBstMidEndSepPunct{\mcitedefaultmidpunct}
{\mcitedefaultendpunct}{\mcitedefaultseppunct}\relax
\EndOfBibitem
\bibitem[Fu \latin{et~al.}(2010)Fu, Santori, Barclay, and
  Beausoleil]{fu2010conversion}
Fu,~K.-M.; Santori,~C.; Barclay,~P.; Beausoleil,~R. Conversion of neutral
  nitrogen-vacancy centers to negatively charged nitrogen-vacancy centers
  through selective oxidation. \emph{Applied Physics Letters} \textbf{2010},
  \emph{96}, 121907\relax
\mciteBstWouldAddEndPuncttrue
\mciteSetBstMidEndSepPunct{\mcitedefaultmidpunct}
{\mcitedefaultendpunct}{\mcitedefaultseppunct}\relax
\EndOfBibitem
\bibitem[Ohno \latin{et~al.}(2012)Ohno, Joseph~Heremans, Bassett, Myers, Toyli,
  Bleszynski~Jayich, Palmstr{\o}m, and Awschalom]{ohno2012engineering}
Ohno,~K.; Joseph~Heremans,~F.; Bassett,~L.~C.; Myers,~B.~A.; Toyli,~D.~M.;
  Bleszynski~Jayich,~A.~C.; Palmstr{\o}m,~C.~J.; Awschalom,~D.~D. Engineering
  shallow spins in diamond with nitrogen delta-doping. \emph{Applied Physics
  Letters} \textbf{2012}, \emph{101}, 082413\relax
\mciteBstWouldAddEndPuncttrue
\mciteSetBstMidEndSepPunct{\mcitedefaultmidpunct}
{\mcitedefaultendpunct}{\mcitedefaultseppunct}\relax
\EndOfBibitem
\bibitem[Fujiwara \latin{et~al.}()Fujiwara, Tsukahara, Sera, Yukawa, Baba,
  Shikata, and Hashimoto]{fujiwara2018tracking}
Fujiwara,~M.; Tsukahara,~R.; Sera,~Y.; Yukawa,~H.; Baba,~Y.; Shikata,~S.;
  Hashimoto,~H. Tracking the spin properties of single nitrogen-vacancy centers
  in nanodiamonds in aqueous buffer solutions. \emph{arXiv:1802.07431.} \relax
\mciteBstWouldAddEndPunctfalse
\mciteSetBstMidEndSepPunct{\mcitedefaultmidpunct}
{}{\mcitedefaultseppunct}\relax
\EndOfBibitem
\bibitem[Fujiwara \latin{et~al.}(2018)Fujiwara, Shikano, Tsukahara, Shikata,
  and Hashimoto]{fujiwara2018observation}
Fujiwara,~M.; Shikano,~Y.; Tsukahara,~R.; Shikata,~S.; Hashimoto,~H.
  Observation of the linewidth broadening of single spins in diamond
  nanoparticles in aqueous fluid and its relation to the rotational Brownian
  motion. \emph{Scientific Reports} \textbf{2018}, \emph{8}, 14773\relax
\mciteBstWouldAddEndPuncttrue
\mciteSetBstMidEndSepPunct{\mcitedefaultmidpunct}
{\mcitedefaultendpunct}{\mcitedefaultseppunct}\relax
\EndOfBibitem
\bibitem[Stanwix \latin{et~al.}(2010)Stanwix, Pham, Maze, Le~Sage, Yeung,
  Cappellaro, Hemmer, Yacoby, Lukin, and Walsworth]{stanwix2010coherence}
Stanwix,~P.~L.; Pham,~L.~M.; Maze,~J.~R.; Le~Sage,~D.; Yeung,~T.~K.;
  Cappellaro,~P.; Hemmer,~P.~R.; Yacoby,~A.; Lukin,~M.~D.; Walsworth,~R.~L.
  Coherence of nitrogen-vacancy electronic spin ensembles in diamond.
  \emph{Physical Review B} \textbf{2010}, \emph{82}, 201201\relax
\mciteBstWouldAddEndPuncttrue
\mciteSetBstMidEndSepPunct{\mcitedefaultmidpunct}
{\mcitedefaultendpunct}{\mcitedefaultseppunct}\relax
\EndOfBibitem
\bibitem[Maze \latin{et~al.}(2008)Maze, Taylor, and Lukin]{maze2008electron}
Maze,~J.; Taylor,~J.; Lukin,~M. Electron spin decoherence of single
  nitrogen-vacancy defects in diamond. \emph{Physical Review B} \textbf{2008},
  \emph{78}, 094303\relax
\mciteBstWouldAddEndPuncttrue
\mciteSetBstMidEndSepPunct{\mcitedefaultmidpunct}
{\mcitedefaultendpunct}{\mcitedefaultseppunct}\relax
\EndOfBibitem
\bibitem[Rondin \latin{et~al.}(2014)Rondin, Tetienne, Hingant, Roch,
  Maletinsky, and Jacques]{rondin2014magnetometry}
Rondin,~L.; Tetienne,~J.; Hingant,~T.; Roch,~J.; Maletinsky,~P.; Jacques,~V.
  Magnetometry with nitrogen-vacancy defects in diamond. \emph{Reports on
  Progress in Physics} \textbf{2014}, \emph{77}, 056503\relax
\mciteBstWouldAddEndPuncttrue
\mciteSetBstMidEndSepPunct{\mcitedefaultmidpunct}
{\mcitedefaultendpunct}{\mcitedefaultseppunct}\relax
\EndOfBibitem
\bibitem[Pham(2013)]{pham2013magnetic}
Pham,~L.~M. Magnetic field sensing with nitrogen-vacancy color centers in
  diamond. Ph.D.\ thesis, Massachusetts Institute of Technology, 2013\relax
\mciteBstWouldAddEndPuncttrue
\mciteSetBstMidEndSepPunct{\mcitedefaultmidpunct}
{\mcitedefaultendpunct}{\mcitedefaultseppunct}\relax
\EndOfBibitem
\bibitem[Takahashi \latin{et~al.}(2008)Takahashi, Hanson, van Tol, Sherwin, and
  Awschalom]{takahashi2008quenching}
Takahashi,~S.; Hanson,~R.; van Tol,~J.; Sherwin,~M.~S.; Awschalom,~D.~D.
  Quenching spin decoherence in diamond through spin bath polarization.
  \emph{Physical Review Letters} \textbf{2008}, \emph{101}, 047601\relax
\mciteBstWouldAddEndPuncttrue
\mciteSetBstMidEndSepPunct{\mcitedefaultmidpunct}
{\mcitedefaultendpunct}{\mcitedefaultseppunct}\relax
\EndOfBibitem
\bibitem[Ishikawa \latin{et~al.}(2012)Ishikawa, Fu, Santori, Acosta,
  Beausoleil, Watanabe, Shikata, and Itoh]{ishikawa2012optical}
Ishikawa,~T.; Fu,~K.-M.~C.; Santori,~C.; Acosta,~V.~M.; Beausoleil,~R.~G.;
  Watanabe,~H.; Shikata,~S.; Itoh,~K.~M. Optical and spin coherence properties
  of nitrogen-vacancy centers placed in a 100 nm thick isotopically purified
  diamond layer. \emph{Nano Letters} \textbf{2012}, \emph{12}, 2083--2087\relax
\mciteBstWouldAddEndPuncttrue
\mciteSetBstMidEndSepPunct{\mcitedefaultmidpunct}
{\mcitedefaultendpunct}{\mcitedefaultseppunct}\relax
\EndOfBibitem
\bibitem[Tetienne \latin{et~al.}(2013)Tetienne, Hingant, Rondin, Cavailles,
  Mayer, Dantelle, Gacoin, Wrachtrup, Roch, and Jacques]{tetienne2013spin}
Tetienne,~J.-P.; Hingant,~T.; Rondin,~L.; Cavailles,~A.; Mayer,~L.;
  Dantelle,~G.; Gacoin,~T.; Wrachtrup,~J.; Roch,~J.-F.; Jacques,~V. Spin
  relaxometry of single nitrogen-vacancy defects in diamond nanocrystals for
  magnetic noise sensing. \emph{Physical Review B} \textbf{2013}, \emph{87},
  235436\relax
\mciteBstWouldAddEndPuncttrue
\mciteSetBstMidEndSepPunct{\mcitedefaultmidpunct}
{\mcitedefaultendpunct}{\mcitedefaultseppunct}\relax
\EndOfBibitem
\bibitem[Myers \latin{et~al.}(2014)Myers, Das, Dartiailh, Ohno, Awschalom, and
  Jayich]{myers2014probing}
Myers,~B.~A.; Das,~A.; Dartiailh,~M.; Ohno,~K.; Awschalom,~D.~D.; Jayich,~A.~B.
  Probing surface noise with depth-calibrated spins in diamond. \emph{Physical
  Review Letters} \textbf{2014}, \emph{113}, 027602\relax
\mciteBstWouldAddEndPuncttrue
\mciteSetBstMidEndSepPunct{\mcitedefaultmidpunct}
{\mcitedefaultendpunct}{\mcitedefaultseppunct}\relax
\EndOfBibitem
\bibitem[Ryan \latin{et~al.}(2018)Ryan, Stacey, O’Donnell, Ohshima, Johnson,
  Hollenberg, Mulvaney, and Simpson]{ryan2018impact}
Ryan,~R.~G.; Stacey,~A.; O’Donnell,~K.~M.; Ohshima,~T.; Johnson,~B.~C.;
  Hollenberg,~L.~C.; Mulvaney,~P.; Simpson,~D.~A. Impact of Surface
  Functionalization on the Quantum Coherence of Nitrogen-Vacancy Centers in
  Nanodiamonds. \emph{ACS Applied Materials \& Interfaces} \textbf{2018},
  \emph{10}, 13143--13149\relax
\mciteBstWouldAddEndPuncttrue
\mciteSetBstMidEndSepPunct{\mcitedefaultmidpunct}
{\mcitedefaultendpunct}{\mcitedefaultseppunct}\relax
\EndOfBibitem
\bibitem[Conner and Schmid(2003)Conner, and Schmid]{conner2003regulated}
Conner,~S.~D.; Schmid,~S.~L. Regulated portals of entry into the cell.
  \emph{Nature} \textbf{2003}, \emph{422}, 37\relax
\mciteBstWouldAddEndPuncttrue
\mciteSetBstMidEndSepPunct{\mcitedefaultmidpunct}
{\mcitedefaultendpunct}{\mcitedefaultseppunct}\relax
\EndOfBibitem
\bibitem[Kim \latin{et~al.}(2014)Kim, Mamin, Sherwood, Rettner, Frommer, and
  Rugar]{kim2014effect}
Kim,~M.; Mamin,~H.; Sherwood,~M.; Rettner,~C.; Frommer,~J.; Rugar,~D. Effect of
  oxygen plasma and thermal oxidation on shallow nitrogen-vacancy centers in
  diamond. \emph{Applied Physics Letters} \textbf{2014}, \emph{105},
  042406\relax
\mciteBstWouldAddEndPuncttrue
\mciteSetBstMidEndSepPunct{\mcitedefaultmidpunct}
{\mcitedefaultendpunct}{\mcitedefaultseppunct}\relax
\EndOfBibitem
\bibitem[Takimoto \latin{et~al.}(2010)Takimoto, Chano, Shimizu, Okabe, Ito,
  Morita, Kimura, Inubushi, and Komatsu]{takimoto2010preparation}
Takimoto,~T.; Chano,~T.; Shimizu,~S.; Okabe,~H.; Ito,~M.; Morita,~M.;
  Kimura,~T.; Inubushi,~T.; Komatsu,~N. Preparation of fluorescent diamond
  nanoparticles stably dispersed under a physiological environment through
  multistep organic transformations. \emph{Chemistry of Materials}
  \textbf{2010}, \emph{22}, 3462--3471\relax
\mciteBstWouldAddEndPuncttrue
\mciteSetBstMidEndSepPunct{\mcitedefaultmidpunct}
{\mcitedefaultendpunct}{\mcitedefaultseppunct}\relax
\EndOfBibitem
\bibitem[Bolshedvorskii \latin{et~al.}(2017)Bolshedvorskii, Vorobyov, Soshenko,
  Shershulin, Javadzade, Zeleneev, Komrakova, Sorokin, Belobrov, Smolyaninov,
  and Akimov]{bolshedvorskii2017single}
Bolshedvorskii,~S.~V.; Vorobyov,~V.~V.; Soshenko,~V.~V.; Shershulin,~V.~A.;
  Javadzade,~J.; Zeleneev,~A.~I.; Komrakova,~S.~A.; Sorokin,~V.~N.;
  Belobrov,~P.~I.; Smolyaninov,~A.~N. \latin{et~al.}  Single bright NV centers
  in aggregates of detonation nanodiamonds. \emph{Optical Materials Express}
  \textbf{2017}, \emph{7}, 4038--4049\relax
\mciteBstWouldAddEndPuncttrue
\mciteSetBstMidEndSepPunct{\mcitedefaultmidpunct}
{\mcitedefaultendpunct}{\mcitedefaultseppunct}\relax
\EndOfBibitem
\end{mcitethebibliography}
\providecommand{\latin}[1]{#1}
\makeatletter
\providecommand{\doi}
  {\begingroup\let\do\@makeother\dospecials
  \catcode`\{=1 \catcode`\}=2\doi@aux}
\providecommand{\doi@aux}[1]{\endgroup\texttt{#1}}
\makeatother
\providecommand*\mcitethebibliography{\thebibliography}
\csname @ifundefined\endcsname{endmcitethebibliography}
  {\let\endmcitethebibliography\endthebibliography}{}

\end{document}